\documentstyle[12pt,epsfig]{article} 
\newcommand{\tr}{\hbox{tr}}
\newcommand{\Tr}{\hbox{Tr}} 

\newcommand{\unit}{\mbox{\bf\sf 1}}
\newcommand{\neu}[1]{\mbox{$#1^{\rm (n)}$}}
\newcommand{\ch}[1]{\mbox{$#1^{\rm (ch)}$}}
\newcommand{\vek}[1]{\mbox{\boldmath$#1$\unboldmath}}
\newcommand{\HH}{\mbox{${\sf H}$}}
\newcommand{\LL}{\mbox{${\sf L}$}}
\newcommand{\DD}{\mbox{${\sf D}$}}
\newcommand{\PP}{\mbox{${\sf P}$}}
\newcommand{\TT}{\mbox{${\sf T}$}}
\newcommand{\MM}{\mbox{${\sf M}$}}
\newcommand{\longit}{\mbox{${\vek{\ell} }$}}
\newcommand{\trans}{\mbox{${\vek{t}}$}}

\newcommand{\dalembert}{\mbox{$\displaystyle\Box$}}

\begin{document} 
\vglue 1truecm
\vbox{ DRAFT UNITU-THEP-13/1997 \hfill \today}
  
\vfil
\centerline{\Large\bf Field Strength Formulation of}
\centerline{\Large\bf $SU(2)$ Yang-Mills Theory }
\centerline{\Large\bf in the Maximal Abelian Gauge:} 
\centerline{\Large\bf Perturbation Theory${}^1$}
  
\bigskip
\centerline{M.\ Quandt${}^2$, H.\ Reinhardt } 
\bigskip
\vspace{1 true cm} 
\centerline{ Institut f\"ur Theoretische Physik, Universit\"at 
   T\"ubingen }
\centerline{D--72076 T\"ubingen, Germany.}
\bigskip
\vskip 1.5cm

\begin{abstract}
\noindent 
We present a reformulation of $SU(2)$ Yang-Mills theory in the maximal
Abe\-lian gauge, where the non-Abelian gauge field components are exactly
integrated out at the expense of a new Abelian tensor field. The latter 
can be treated in a semiclassical approximation and the corresponding
saddle point equation is derived. Besides the non-trivial solutions, 
which are presumably related to non-perturbative interactions for the
Abelian gauge field, the equation of motion for the tensor fields allows 
for a trivial solution as well. We show that the semiclassical expansion 
around this trivial solution is equivalent to the standard perturbation 
theory. In particular, we calculate the one-loop $\beta$-function for the 
running coupling constant in this approach and reproduce the standard
result.

\end{abstract}

\vfil
\hrule width 5truecm
\vskip .2truecm
\begin{quote} 
$^1$ Supported in part by DFG under contract Re 856/1-3.\\
$^2$ Supported by ``Graduiertenkolleg: Hadronen und Kerne''.
\end{quote}
\eject
\section{ Introduction } 
\label{sec:1}

QCD is generally accepted as the theory of strong interactions. It has been
successfully tested in the high energy regime, where perturbation theory is
applicable. On the other hand the low energy sector of the theory, and in
particular the confinement mechanism is not well understood. A theoretically 
appealing mechanism for confinement has been proposed some time ago by 
{}'t Hooft and Mandelstam, who conjectured that confinement is realised by the 
dual Meissner effect \cite{R1}. This confinement scenario assumes that the QCD
ground state consist of a condensate of magnetic monopoles (dual superconductor),
which squeezes the colour electric field of colour charges into flux tubes.

Magnetic monopoles arise in Yang-Mills theory in the so called Abelian gauges,
which were proposed by {}'t Hooft \cite{R2}. Although in these gauges the 
magnetic monopoles arise as gauge artifacts, they may nevertheless represent 
the dominant infrared degrees of freedom in these gauges. The realisation of 
the dual Meissner effect in the Abelian gauges has recently received strong 
support from lattice calculations \cite{R3}. These lattice calculations also 
show that there is a preferred Abelian gauge, the so-called maximal Abelian 
gauge. In this gauge one, in fact, observes Abelian dominance and furthermore 
dominance of magnetic monopoles, i.~e.~about 92 \% of the full string tension 
comes from the Abelian field configurations and furthermore 95 \% of the 
Abelian string tension is reproduced by the magnetic monopoles alone \cite{R4}.
The Abelian dominance indicates that the low energy sector of QCD can probably 
most efficiently be described in terms of an effective Abelian 
theory\footnote{This Abelian theory should account for the non-Abelian 
sampling of field configurations. Let us emphasise that the Abelian dominance 
is observed with a non-Abelian sampling of field configurations, i.~e.~the 
field configurations are weighted by the full non-Abelian Yang-Mills action.}. 
Unfortunately the underlying effective Abelian theory of QCD, which gives rise 
to the dual Meissner effect, is not known. The dual Meissner effect also seems 
to be the confinement mechanism in supersymmetric gauge theories \cite{R5}.

Apparently the dual Meissner effect can be most efficiently described in a dual
formulation, which is known to exist for quantum electrodynamics. The
transition to the dual theory basically amounts to an interchange of electric
and magnetic fields, and moreover to an inversion of the coupling constant. 

Unfortunately in the strict sense the dual theory of non-supersymmetric 
Yang-Mills theory is not known and does perhaps not exist. However, there 
exists a non-local formulation in terms of "dual potentials" which are 
introduced as functional Fourier conjugates to the Bianchi forms \cite{halp}. 
Furthermore, within loop space formulation of gauge theory an extended duality 
transformation for non-Abelian theories has been introduced, which for 
electrodynamics reduces to the standard duality transformation \cite{R6}. 
Although this extended dual formulation has some attractive features as, for 
example, it allows to construct explicitly {}'t Hooft's magnetic disorder 
operator \cite{R7}, it seems to be rather inconvenient for calculational 
purposes.

There have been also attempts to construct the dual theory of QCD
phenomenologically \cite{R8}. Furthermore, there exist a microscopic approach 
to the dual description of QCD, which has not been fully appreciated in the 
past. This is the so-called field strength approach \cite{R9}, which 
reformulates the Yang-Mills theory in terms of the field strength tensor and 
comes very close to a dual description as has been recently discussed
\cite{R10}.

In the present paper by reformulating the field strength approach in the
maximal Abelian gauge, we derive an effective Abelian theory.\footnote{Some
preliminary results of this approach have been reported in ref.~\cite{R10a}.} 
For simplicity we will confine ourself to the gauge group $SU(2)$. In this case
we succeed to exactly integrate out the non-Abelian part of the gauge field at 
the expense of the introduction of an Abelian tensor field. As a first 
illustration of the emerging effective Abelian theory we calculate the 
one-loop $\beta$-function and reproduce the standard Yang-Mills result.

\medskip

Let us also mention that recently the field strength approach has been
reconsidered from the point of view of the so-called BF-theory \cite{R11}. 
In this context Yang-Mills theory can be regarded as a deformation of the 
BF-theory. The corresponding Feynman rules and one loop $\beta$-function were 
considered in ref.~\cite{R12} in the Lorentz gauge, making use of special 
properties of this gauge, like the finiteness of the gluon-ghost vertex, which 
is absent in the nonlinear maximal Abelian gauge. 
In the present paper we will perform a one-loop calculation in the 
maximal Abelian gauge in a modified field strength approach, thereby showing 
explicitly that vertex corrections and contributions from the charged gluon 
propagator cancel on one-loop level. This cancellation is similar 
to the background field calculation, where only the propagator of the 
background field contributes to the $\beta$-function. In fact, in our case the 
Abelian gluon field figures in this context as a background field.

The organisation of the paper is as follows:
After fixing our conventions, we discuss the Faddeev-Popov gauge fixing
for the Maximal Abelian gauge. By introducing an Abelian tensor field
$\chi_{\mu\nu}$ similar to the field strength approach \cite{R9}, we can 
exactly integrate out the charged gauge field components, resulting in an 
effective Abelian theory given implicitly as an integral over the tensor field
$\chi$ and the FP ghosts. We also investigate the general form of the 
resulting charged gluon propagator. By integrating out the charged gluon fields
a good deal of quantum fluctuations has been included and the remaining
integral over the tensor field $\chi$ can presumably be treated in 
saddle point approximation.
In section \ref{sec:2.3}, we derive the corresponding equation of motion for 
$\chi$ and show that, unlike the standard field strength approach, a trivial 
solution $\chi=0$ also exists. 

In section \ref{sec:3} we examine the nature of the semiclassical expansion
around this trivial saddle point. As expected, it corresponds to the 
standard perturbative expansion. After deriving an explicit form of the 
effective action to one-loop order, we show in section \ref{sec:3.2} that
there is a cancellation between the charged field propagators and the
vertex correction, so that the $\beta$-function receives contributions only 
from the neutral gluon vacuum polarisation, a state of affairs familiar from
the background gauge formalism. 
Finally, we complete the calculation of the one-loop
$\beta$-function by evaluating the corrections to the neutral gluon propagator.
The result coincides with the known expression.

We conclude in section \ref{sec:4} with an outlook on possible non-perturbative
applications of our approach.  

\section{The Effective Abelian Theory} 
 \label{sec:2}
We consider pure $SU(N)$ Yang-Mills theory on an Euclidean four-manifold 
$\cal M$ with the action given by
\begin{equation}
S_{\rm YM}[A] = \frac{1}{4g^2}\int\limits_{\cal M}\,d^4 x\,
F^a_{\mu\nu}(x) F_a^{\mu\nu}(x).
\label{<1>}
\end{equation}
Here, $F_{\mu\nu} = F_{\mu\nu}^a T^a= \partial_\mu A_\nu - \partial_\nu A_\mu 
+ [A_\mu,A_\nu]$ denotes the field strength tensor of the gauge
field $A_\mu \equiv A_\mu^a T^a$, and the generators 
$T^a\,\,(a=1,\ldots,N^2-1)$ of the Lie-algebra 
$[T^a,T^b] = f^{a b c} T^c$
are taken to be anti-hermitian and normalised according to
\begin{equation}
\tr\left(T^a T^b\right) = -\frac{1}{2} \delta^{a b}.
\label{<2>}
\end{equation}
For the case of $SU(2)$, $T^a = \sigma^a/2 i$ where $\sigma^a$ are the
Pauli matrices.  Under a gauge transformation $U = \exp(-\omega^a T^a)$,
\begin{equation}
A_\mu\stackrel{U}{\longrightarrow} A_\mu^U := UA_\mu U^\dagger + U\partial_\mu 
U^\dagger .
\label{<4>}
\end{equation}
For infinitesimal $U\approx\unit-\omega^a T^a$, this means
\[
\delta A_\mu^a = f^{abc}
A_\mu^b \,\omega^c + \partial_\mu\,\omega^a\equiv \hat{D}_\mu^{ab} \omega^b, 
\]
where we also introduced the covariant derivative
\begin{equation}
D_\mu = \partial_\mu + A_\mu \quad ;\quad \hat{D}_\mu^{ab} =
\partial_\mu \delta^{ab} - f^{abc} A_\mu^c \equiv \partial_\mu \delta^{ab}
+ \hat{A}_\mu^{ab}.
\label{<5>} 
\end{equation}
Finally, quantities with a hat ($\,\hat{}\,$) denote Lie-algebra elements in 
the adjoint representation, with generators given by 
$\left[\hat{T}^a\right]_{bc} = - f_{abc}$. 

\subsection{Maximal Abelian Gauge}
\label{sec:2.1}
The basic idea of Abelian gauges \cite{R2} is to remove as many non-Abelian
degrees of freedom as possible by partially fixing the gauge, leaving 
a theory with a residual Abelian gauge symmetry. Technically, this is 
accomplished by the so-called Cartan decomposition.

Every Lie algebra has a largest Abelian subalgebra spanned by 
a maximal set of commuting generators.
Throughout this paper, colour indices $a_0$, $b_0$ etc.~denote the generators 
of this Cartan subalgebra, while letters with a bar, $\bar{a}$, $\bar{b}$ 
etc.~are  reserved for the remaining generators. For the simple Lie algebras 
$su(N)$ we can always adjust the Cartan decomposition in such a way that the 
Abelian generators $T^{a_0}$ are diagonal, and the remaining $T^{\bar{a}}$ have 
vanishing diagonal elements. 

Since the generators of the Cartan algebra commute, their picture under the
exponential is an Abelian subgroup $H$ of the corresponding Lie group $G$, and 
one finds an analogous decomposition for the group
\begin{equation}
G = H \otimes G/H \quad;\quad
H \equiv \left\{\exp\left(-\omega_{a_0} T^{a_0}\right)\right\},
\label{<8>}
\end{equation}
where the quotient $G/H$ is called coset. For $SU(N)$, there are $N-1$
commuting generators (the rank of the group), and the Cartan subgroup $H$ is a 
(reducible) representation of $U(1)^{N-1}$. Our goal is to 
reduce the gauge symmetry to this $U(1)^{N-1}$ by fixing the coset.

In the following, we will mainly consider the case $SU(N=2)$, for which the
Cartan decomposition of the gauge field is
\begin{equation}
A_\mu(x) = a_\mu(x)\,T^3 + \sum\limits_{\bar{a}=1}^{2} 
A_\mu^{\bar{a}}(x)\,T^{\bar{a}} \equiv \neu{A_\mu}(x) + \ch{A_\mu}(x).
\label{<9>}
\end{equation}
The superscripts ${}^{\rm (ch)}$ and ${}^{\rm (n)}$ for "charged" and "neutral"
refer to the transformation properties under the residual Abelian gauge
group $\Omega = e^{-\omega T^3} \in H$:
\begin{equation}
\ch{A_\mu} \longrightarrow \Omega\cdot\ch{A_\mu} \cdot\Omega^\dagger 
\quad ; \quad
a_\mu \longrightarrow a_\mu + \partial_\mu \omega .
\label{<10>}
\end{equation}
Under this residual $U(1)$, $A_\mu^{\rm (ch)}$ transforms as a charged matter 
field in the adjoint representation, while the diagonal part 
$A_\mu^{\rm (n)} = a_\mu \, T^3$ acts like a photon. Note that the symmetry
(\ref{<10>}) involves the diagonal part of the {\em original} gauge group and 
gluon field. Therefore, it constitutes an {\em electric} $U(1)$.

To fix the coset we need $N(N-1) = 2$ conditions, which are invariant under
the Cartan group $U(1)$. Several propositions have been made since {}'t Hooft's
original work \cite{R2}, but in this paper, for the reasons mentioned in
the introduction, we will use the so-called 
{\em maximal Abelian gauge} ({\sf MAG}):
\begin{equation}
\left[D_\mu^{\rm (n)}, \ch{A_\mu}\right] = \partial_\mu \ch{A_\mu} + 
\left[\neu{A_\mu},\ch{A_\mu} \right] \equiv 0 .
\label{<11>}
\end{equation} 
The main motivation for the use of Abelian gauges is, besides the 
appearance of magnetic monopoles, that it should facilitate integrating
out the charged gauge field components $\ch{A_\mu}$  leaving an effective 
Abelian theory. The quantisation of the latter still requires a gauge fixing 
for the neutral photon, which in the present paper will be done by the usual
Lorentz condition $\partial_\mu A_\mu^{\rm (n)} \equiv 0$. The complete 
gauge fixing constraints thus read 
\begin{equation}
\left[D_\mu^{\rm (n)}, A_\mu\right] = \partial_\mu A_\mu + 
\left[\neu{A_\mu},A_\mu \right] \equiv 0.
\label{<12>}
\end{equation}
The standard Faddeev-Popov (or BRST) quantisation of the gauge (\ref{<12>})
leads to the following gauge-fixing insertion in the path integral
\begin{equation}
\delta_{\rm gf} = \delta[\partial_\mu a_\mu] \,\det\left(-\dalembert\right)
\cdot\delta\left[\neu{D_\mu} \ch{A_\mu}\right]\,\det {\sf M}[a,A].
\label{<12a>}
\end{equation}
The splitting of the FP determinant in an Abelian and charged part only holds
upon using the {\sf MAG}-condition (\ref{<11>}), i.e.~the second 
delta-function in $\delta_{\rm gf}$ has to be implemented {\em exactly}. We 
may, however, relax the Abelian Lorentz-condition in the usual way by 
introducing a gauge-fixing term in the action:
\begin{equation}
\delta_{\rm gf} = \exp\left(-\frac{1}{2\hbar g^2 \alpha}\int (\partial \cdot
a)^2\right)\det\left(-\dalembert\right)
\cdot\delta\left[\neu{D_\mu} \ch{A_\mu}\right]\,\det {\sf M}[a,A]
\label{<12b>}
\end{equation}
where $\alpha$ is a gauge fixing parameter. Note that the first two factors
in (\ref{<12b>}) represent the standard FP insertion for the Abelian
Lorentz gauge, so that the last two terms may be considered as implementation
of the {\sf MAG} (\ref{<11>}) alone.

In the $SU(2)$ case, we have 
\begin{equation}
{\sf M}^{\bar{a}\bar{b}} := - \left[\neu{D_\mu} \neu{D_\mu}\right]^{\bar{a}
\bar{b}} + \left(A_\mu^{\bar{a}}A_\mu^{\bar{b}} - 
A_\mu^{\bar{c}}A_\mu^{\bar{c}} \delta^{\bar{a}\bar{b}}\right)
\label{<13>}
\end{equation}
Introducing charged ghosts and a multiplier field $\ch{\phi}$ to express the
{\sf MAG} delta-function in (\ref{<12b>}) by its Fourier representation,
we can rewrite the generating functional for $SU(2)$ YM-theory in the
following form:
\begin{eqnarray}
Z[j] &=& \int {\cal D} (\ch{A},a,\eta,\bar{\eta},\phi)\,
\exp\left\{-\frac{1}{\hbar}S_{\rm eff}+ 
\int \left(j_{\mu}^{\bar{a}} A_{\mu}^{\bar{a}} + 
j_\mu^3 a_\mu \right)\,d^4x \right\} 
\label{<14>}\\
S_{\rm eff} &=& S_{\rm YM}[a,\ch{A}] + \frac{1}{2 \alpha g^2} \int 
\left(\partial\cdot a\right)^2 + 
\frac{1}{g^2} \int \bar{\eta}^{\bar{a}}
\MM^{\bar{a}\bar{b}} \eta^{\bar{b}} + \frac{i}{g^2}  \int \phi^{\bar{a}}
\neu{D_{\mu,\bar{a}\bar{b}}} A_\mu^{\bar{b}}.
\nonumber
\end{eqnarray}
For simplicity, we will from now on use the notation
$a_\mu \equiv A_\mu^3$ for the neutral gluon (photon) in $SU(2)$. 

\subsection{Derivation of the Effective Abelian Theory}
\label{sec:2.2}
In principle, we can map $SU(2)$ YM into an effective Abelian theory by
integrating out the charged field components $A^{\rm (ch)}$ from the 
generating functional (\ref{<14>}). Unfortunately, the YM action
is quartic in the charged gauge field components, so that the integration 
cannot be performed exactly. 
Furthermore, the singularities of the gauge fixing procedure will not only 
lead to monopoles in the neutral photon, but will also induce non-trivial 
topology in $\ch{A}$ \cite{R13}. 

To make the dependence on $\ch{A}$ more explicit, we decompose the YM field
strength in charged and neutral parts:
\begin{eqnarray}
F_{\mu\nu} &=& \left(\partial_\mu \neu{A_\nu} - \partial_\nu \neu{A_\mu}\right)
+ \left(\left[\neu{D_\mu},\ch{A_\nu}\right] - \left[\neu{D_\nu},\ch{A_\mu}
\right]\right) +\left[\ch{A_\mu},\ch{A_\nu}\right] \nonumber \\
&\equiv& f_{\mu\nu} + V_{\mu\nu} + T_{\mu\nu} .
\label{<15>}
\end{eqnarray}
For an arbitrary gauge group $G=SU(N)$, the Abelian field strength 
$f_{\mu\nu}$ is always neutral, i.e.~it lives in
the Cartan subalgebra, while $V_{\mu\nu}$ is always
charged. The problematic term is the commutator 
$T_{\mu\nu} = \left[\ch{A_\mu},\ch{A_\nu}\right]$, which in general
contains both types of generators. However, in the special case $G=SU(2)$
$T_{\mu\nu}$ only lives in the Cartan subalgebra. 
From the orthogonal normalisation (\ref{<2>}) of the generators, this entails
\begin{equation}
\tr \left(f_{\mu\nu}\cdot V_{\mu\nu}\right) = 
\tr \left(T_{\mu\nu}\cdot V_{\mu\nu}\right) \equiv 0.
\label{<16>}
\end{equation}
In other words, this expresses the absence of a charged three gluon vertex in
the YM-action for the case of only two colours.
From now on, we will therefore focus on the gauge group
$SU(2)$. 
Using the decomposition (\ref{<15>}) in the YM action, we find
\begin{equation}
S_{\rm YM} = -\frac{1}{2g^2} \int \tr\left\{f_{\mu\nu}f^{\mu\nu} + 
V_{\mu\nu}V^{\mu\nu} + T_{\mu\nu}T^{\mu\nu} + 2 f_{\mu\nu}T^{\mu\nu}
\right\}\,d^4x . 
\label{<17>}
\end{equation}
The $V^2$-term can be simplified by a partial integration, using the
the {\sf MAG}-condition:
\begin{eqnarray}
\int \tr\left(V_{\mu\nu} V^{\mu\nu}\right)\,d^4x &=& 
\int A_\mu^{\bar{a}} \left[\neu{\hat{D}_\sigma}\neu{\hat{D}_\sigma}
\right]_{\bar{a}\bar{b}}A_\mu^{\bar{b}}\,d^4x +
2\int \tr\left(T_{\mu\nu} f^{\mu\nu}\right)\,d^4x + \mbox{surface}
\nonumber \\ 
\mbox{surface}&=& 2 \oint \tr\left(V_{\mu\nu}\cdot\ch{A_\nu}\right)\,
d\sigma^\mu . 
\label{<18>}
\end{eqnarray}
This allows to reformulate the YM action as
\begin{eqnarray}
S_{\rm YM} &=& \frac{1}{4g^2} \int  \left(f^3_{\mu\nu}\right)^2\,d^4 x + 
\frac{1}{2g^2} \int A_\mu^{\bar{a}}\left[ - \neu{\hat{D}_\sigma} 
\neu{\hat{D}_\sigma}\delta_{\mu\nu}  - 2 \hat{f}_{\mu\nu}
\right]_{\bar{a}\bar{b}} A_\nu^{\bar{b}} +
\nonumber \\
& & + \frac{1}{4g^2} \int \left(T^3_{\mu\nu}\right)^2\,d^4x
+ {\rm surface} .
\label{<19>}
\end{eqnarray}
The usual approach of treating the $(T_{\mu\nu})^2$ term (which is  
quartic in the charged gauge field)
perturbativelyt is probably not appropriate in the low energy
regime, where the induced self-interaction of the Abelian field $a_\mu$
is expected to be strong in order to trigger monopole
condensation. Therefore, some non-perturbative treatment must be adopted, 
and the structure of the action suggests a path integral linearisation.

For this purpose, let us introduce a neutral, antisymmetric tensor field 
$\neu{\chi_{\mu\nu}} = \chi_{\mu\nu} T^3$ by the 
identity\footnote{Sums over repeated Lorentz indices
run from $1\ldots 4$, but the path integration is only performed over the
{\em independent} components of $\chi$, i.e.~${\cal D}\chi = 
\prod\limits_{\mu < \nu} {\cal D} \chi_{\mu\nu}(x)$.}
\begin{eqnarray}
\exp\left\{-\frac{1}{4\hbar g^2} \int \left(T^3_{\mu\nu}\right)^2\,d^4x
\right\} = \int {\cal D} \chi_{\mu\nu} \exp\Bigg\{\!\!\!\!\!&-&\!\!\!\!
\frac{1}{4\hbar g^2}
\int \left(\chi_{\mu\nu}\right)^2\,d^4x + \nonumber\\
{}\!\!\!\!&+&\!\!\!\!\frac{i}{2\hbar g^2}\int
T^3_{\mu\nu} \chi_{\mu\nu}\,d^4x\Bigg\}.
\label{<20>}
\end{eqnarray}
Upon inserting (\ref{<20>}) and (\ref{<19>}) in the generating functional
(\ref{<14>}), we can formally express the effective Abelian theory as 
\begin{equation}
Z[j] = \int {\cal D}a_\mu\,\exp\left\{-\frac{1}{\hbar}\left(
S_0[a] + S_{\rm int}[a,\ch{j}] \right) + \int \neu{j_\mu} a_\mu\,d^4x\right\}.
\label{<22>}
\end{equation}
Here $S_0$ is the standard free Maxwell action
\begin{eqnarray}
S_0 &=& \frac{1}{4 g^2}\int f_{\mu\nu} f_{\mu\nu}\,d^4x + \frac{1}{2 g^2
\alpha}\int \left(\partial_\mu a_\mu\right)^2\,d^4x \nonumber
\\
&=& \frac{1}{2 g^2} \int a_\mu \left[D_0^{-1}\right]_{\mu\nu} a_\nu\,d^4x 
+ \mbox{surface}  
\label{<xx>}
\end{eqnarray}
which defines the tree-level photon propagator
\begin{equation}
\left[D_0\right]_{\mu\nu}(x,y) = \int \frac{d^4 p}{(2\pi)^4}\,e^{-i p(x-y)}\,
\left[\delta_{\mu\nu} + (\alpha-1) 
\frac{p_\mu p_\nu}{p^2}\right]\,\frac{1}{p^2}.
\end{equation}
The remaining term $S_{\rm int}$ describes the self-coupling of the neutral 
gluons (photons) as well as their coupling to the external charged currents. 
Both are non-Abelian effects mediated by the exchange of charged gluons, 
ghosts and multipliers enforcing the gauge constraints. In order to make the 
following formulae more transparent we will mostly suppress colour, Lorentz 
and spacetime indices; a bracket $\langle\cdots\rangle$ symbolises summation 
or integration over all relevant indices. Neglecting surface terms, 
$S_{\rm int}$ is then formally given by
\begin{eqnarray}
e^{-\frac{1}{\hbar} S_{\rm int}} &=& \int{\cal D}
(\ch{A},\chi,\eta,\bar{\eta},\phi)
\exp\Bigg\{-\frac{1}{2\hbar g^2} \left\langle\ch{A}\cdot\HH\cdot\ch{A}
\right\rangle - 
\frac{1}{4 \hbar g^2}\left\langle \chi\cdot\chi\right\rangle - 
\nonumber \\
{} && - \frac{1}{\hbar g^2} 
\left\langle\bar{\eta}\cdot\left(-\neu{\hat{D}}\neu{\hat{D}}\right)\cdot\eta
\right\rangle 
+ \left\langle\ch{A}\cdot\left(\ch{j} - \frac{i}{\hbar g^2} \neu{\hat{D}}\phi
\right)\right\rangle + 
\nonumber \\
{} & & + \left\langle\bar{\xi}\cdot\eta + \bar{\eta}\cdot\xi\right\rangle + 
\left\langle\phi\cdot\varphi\right\rangle\Bigg\}.
\label{<23>}
\end{eqnarray}
For later convenience, we have introduced sources $\eta^{\bar{a}}$ and 
$\varphi^{\bar{a}}$ for the charged ghost $\eta,\bar{\eta}$ and multiplier 
fields $\phi$, respectively. Furthermore, the symmetric operator
\begin{equation}
\HH_{\mu\nu}^{\bar{a}\bar{b}} := - \delta_{\mu\nu}\left(\left[\neu{D_\sigma}
\neu{D_\sigma}\right]^{\bar{a}\bar{b}} + 2 \delta^{\bar{a}\bar{b}} 
\bar{\eta}^{\bar{c}} \eta^{\bar{c}} - 2 \bar{\eta}^{\bar{a}} 
\eta^{\bar{b}}\right) - 2 \hat{f}_{\mu\nu}^{\bar{a}\bar{b}} + 
i \hat{\chi}_{\mu\nu}^{\bar{a}\bar{b}} .
\label{<24>}
\end{equation}
is the inverse of the charged gluon propagator {\em before} gauge fixing to 
the {\sf MAG}. 
Note that $\HH = \HH_0 + \delta\HH$, where the zero field expression
$\HH_0 = -\dalembert$ and its inverse are well defined. In the present
paper, we will use $\HH$ and its inverse only in the perturbative sense, 
i.e.~$\HH^{-1} = \HH_0^{-1} - \HH_0^{-1}\cdot \delta\HH\cdot\HH_0^{-1} + 
\cdots$ which is well defined to any finite order. 
\medskip

At this point, all of the integrations in (\ref{<23>}) should be performed in 
order to obtain the effective photon action. Of course, this cannot be 
done exactly for {\em all} fields in (\ref{<23>}), but at least for 
$\phi$ and $\ch{A}$, the integration can be straightforwardly performed. 
While this is not surprising\footnote{The aim of the path integral 
linearisation is exactly to make this integration feasible at the expense of a 
new auxiliary field $\chi$.} for $\ch{A}$, the exact integral over $\phi$ is 
a very pleasant feature. After all, the factorisation (\ref{<12a>}) of the FP 
determinant requires an exact implementation of the {\sf MAG} condition, 
i.e.~any approximation in $\phi$ would couple charged and neutral 
entries of the FP determinant.\footnote{A gauge fixing term for the {\sf MAG}
would have the same effect. This is the main reason for our use of the 
multiplier field $\phi$.}
In view of the dual superconductor picture,
however, the factorisation (\ref{<12a>}) is crucial as it makes the residual
Abelian $U(1)$ symmetry and its (Lorentz) gauge fixing explicit.
\medskip

Let us have a closer look at the residual {\em electric} $U(1)$ symmetry. 
Under such a diagonal gauge transformation $\omega = e^{-\theta T^3}$ we have
\begin{eqnarray}
a_\mu &\stackrel{\omega}{\longrightarrow}& a_\mu + \omega\cdot\partial_\mu 
\omega^\dagger = a_\mu + \partial_\mu \theta \,\, ,\nonumber \\
\ch{j_\mu}&\stackrel{\omega}{\longrightarrow}& \omega \cdot \ch{j_\mu}
\cdot \omega^\dagger\,\, , \nonumber \\
\ch{A_\mu}&\stackrel{\omega}{\longrightarrow}& \omega \cdot \ch{A_\mu}
\cdot \omega^\dagger\,\, ,\label{<24a>} \\
\chi_{\mu\nu}  &\stackrel{\omega}{\longrightarrow}& \chi_{\mu\nu} \nonumber
\end{eqnarray}
and all the charged fields ($\eta,\bar{\eta},\phi$) and their currents 
transform similar to $\ch{A}$ and $\ch{j}$, respectively. Note that the 
new Abelian tensor field $\chi_{\mu\nu}$ has to be regarded as an 
{\em invariant} with respect to the residual $U(1)$. This is clear
from the introduction (\ref{<20>}) of $\chi_{\mu\nu}$, which shows that it
only couples to the commutator $[\ch{A_\mu},\ch{A_\nu}]$, which in turn is 
invariant under Abelian gauge transformations.

The $U(1)$ symmetry (\ref{<24a>}) is only broken by the Lorentz gauge fixing 
in (\ref{<12b>}), i.e.~by the gauge fixing term in the Maxwell action $S_0$
(\ref{<xx>}).
Furthermore, the integration measure in (\ref{<23>}) is clearly invariant under
the simple gauge rotation (\ref{<24a>}) of the charged fields. Some of the
immediate consequences of this observation are:
\begin{itemize}
\item The complicated interaction term $S_{\rm int}$ from (\ref{<23>}) is
      invariant under the residual Abelian symmetry (\ref{<24a>}) for the
      photon and the charged currents.
\item The same is true for the exponent in (\ref{<23>}) (as can be 
      easily checked by inspection). 
\item Green's functions from the effective theory (\ref{<22>}) 
      obey (Abelian) Ward identities. 
      This has important consequences for the renormalisation in the next 
      chapter.     
\end{itemize}

Returning to the formula (\ref{<23>}), we can now perform the integration
over $\ch{A}$. The resulting exponent is at most quadratic in the multiplier 
field $\phi$, so that an {\em exact} implementation of the {\sf MAG} is 
possible by performing the Gaussian integration over $\phi$. We end up 
with
\begin{eqnarray}
S_{\rm int} &=& -\hbar \int {\cal D}(\chi,\eta,\bar{\eta})
\exp\Bigg\{-\frac{1}{\hbar} {\sf S}[a,\chi,\eta,\bar{\eta}] + \frac{\hbar g^2}
{2}\left\langle\ch{j}\cdot\DD\cdot\ch{j}\right\rangle +
\nonumber\\
{} && + \frac{\hbar g^2}{2} 
\left\langle\varphi\cdot\LL^{-1}\cdot\varphi\right\rangle + 
i \hbar g^2 \left\langle\varphi\cdot\LL^{-1}\neu{\hat{D}_\mu}\HH^{-1}\cdot
\ch{j}\right\rangle + \left\langle\bar{\xi}\eta + 
\bar{\eta}\xi\right\rangle\Bigg\} \label{<25>}
\end{eqnarray}
where
\begin{equation}
{\sf S}[a,\chi,\eta,\bar{\eta}] = \frac{1}{4 g^2} \int \chi^2 + 
\frac{1}{g^2}\int \bar{\eta}\left(-\neu{\hat{D}_\sigma}\neu{\hat{D}_\sigma}
\right) \eta + \frac{\hbar}{2} \Tr \ln \HH + \frac{\hbar}{2} \Tr \ln \LL 
\label{<26>}
\end{equation}
is an effective action. Here we have introduced the definitions
\begin{eqnarray}
\LL_{\bar{a}\bar{b}} &:=&  - \left[\neu{\hat{D}}
\HH^{-1}\neu{\hat{D}}\right]_{\bar{a}\bar{b}}
\nonumber
\\
\DD^{\mu\nu}_{\bar{a}\bar{b}} &:=& 
\left[\HH^{-1} + \left(\HH^{-1} \neu{\hat{D}}\right)\cdot\LL^{-1}\cdot
\left(\neu{\hat{D}}\HH^{-1}\right)\right]^{\mu\nu}_{\bar{a}\bar{b}} .
\label{<28>}
\end{eqnarray}
To prevent the equations from becoming cluttered we have used here a shorthand
notation suppressing the intermediate indices which are summed over. The 
detailed definition of these quantities (with the summation indices restored) 
can be found in appendix \ref{app:1}. Let us stress that $S_{\rm int}$ 
(\ref{<25>}) and in particular the effective action ${\sf S}[a,\chi,\eta,
\bar{\eta}]$ from (\ref{<26>}) is {\em invariant} under the Abelian gauge 
transformation (\ref{<24a>}) provided the employed regularisation prescription
does not spoil this symmetry.
From eq.~(\ref{<25>}) we read off 
that $\DD$ and $\LL^{-1}$ are the propagators of the charged gluons $\ch{A}$
and the multiplier field $\phi$, respectively.\footnote{The remarks given above
on the perturbative existence of $\HH$ and its inverse apply to 
the propagators $\DD$ and $\LL^{-1}$ as well.}

Some comments on these propagators are in order. At zero fields, we find
$\LL_0^{-1} = \unit$ and $\DD_0 = -\dalembert^{-1} \PP_T$, 
where $\PP_T$ denotes the usual transversal projector. Both results are, 
of course, expected. The local form of $\LL_0^{-1}$ simply reflects the fact 
that the field $\phi$ really is a multiplier, not a propagating quantum field.
At zero Abelian field, on the other hand, the {\sf MAG} coincides with the
Lorentz gauge for $\ch{A}$ which yields the transversal propagator $\DD_0$
when implemented exactly.

Prior to integrating out $\phi$ in (\ref{<23>}), the charged gluon propagator
defined by the term quadratic in $\ch{A}$ would be simply $\HH^{-1}$. 
The subsequent integration over $\phi$ implements the {\sf MAG} exactly and
converts the charged gluon propagator from $\HH^{-1}$ to $\DD$.
Therefore, we expect the complicated
form (\ref{<28>}) for $\DD$ (at non-zero fields) to be some kind of 
{\em projection} of $\HH^{-1}$ to the {\sf MAG} subspace. Indeed, we can 
rewrite $\DD$, using the above introduced compact notation, as 
\begin{eqnarray}
\DD = \TT^\dagger \HH^{-1} = \HH^{-1} \TT = \TT^\dagger \HH^{-1} \TT 
\qquad\mbox{with}\qquad
\TT &\equiv& \unit + \neu{D}\,\LL^{-1}\,\neu{D}\,\HH^{-1}
\nonumber \\
\TT^\dagger &\equiv& \unit + \HH^{-1}\,\neu{D}\,\LL^{-1}\,\neu{D} 
\label{<30>}
\end{eqnarray}
As expected, the operator $\TT$ is a projector, i.e. $\TT^2 = \TT$, and 
represents the identity in the space of charged gluon field configurations 
satisfying the ${\sf MAG}$.

The above discussion can be made more transparent by noting that the
{\sf MAG} can be considered as a generalised Lorentz gauge replacing the 
ordinary derivative $\partial_\mu$ by the covariant one $\neu{\hat{D}_\mu}$.
In the same way as the Lorentz gauge eliminates the longitudinal  
component of the gauge field, the {\sf MAG} eliminates the generalised 
longitudinal charged components $\neu{\hat{D}} \ch{A_L} = \longit \ch{A} = 0$,
where the covariant longitudinal and transversal projectors are defined by
\begin{equation}
\longit^{ab}_{\mu\nu} = \left(\neu{\hat{D}_\mu} \frac{1}{\neu{\hat{D}}
\neu{\hat{D}}}\neu{\hat{D}_\nu}\right)^{ab} \quad ; \quad
\trans^{ab}_{\mu\nu} = \delta^{ab}\delta_{\mu\nu} - \longit^{ab}_{\mu\nu}.
\label{<30a>}
\end{equation}
Note also that in the colour subspace of the Cartan algebra these 
operators reduce to the familiar longitudinal and transversal projectors.
It is now easy to see that the propagator (\ref{<28>}) for the charged gluon
is in fact purely {\em transversal\/} in the generalised sense
\begin{equation}
\longit\cdot\DD = 0 \quad , \quad \DD \cdot \longit = 0.
\end{equation}
One could have certainly expected this result, but it is gratifying that
it comes out in our approach without being explicitly implemented.

Finally we can use the generalised projectors (\ref{<30a>}) to simplify 
somewhat the above obtained effective action (\ref{<26>}). From the definition 
(\ref{<28>}) one immediately derives the relation
\begin{equation}
\Tr_{(c)}\ln\LL = \Tr_{(c)}\ln\left(-\neu{\hat{D}_\mu} \neu{\hat{D}_\mu}
\right) + \Tr_{(c,L)}\left(\longit\ln\HH^{-1}\right) + 
\Tr_{(c,L)}\left(\longit\ln\longit\right).
\label{<30b>}
\end{equation}
Here we have explicitly indicated over which discrete indices the trace has
to be taken (where "c" and "L" stand for the colour and Lorentz indices,
respectively). The last two terms in the effective action (\ref{<26>}) can 
then be combined to
\begin{equation}
\Tr_{(c)}\ln\LL+\Tr_{(c,L)}\ln\HH = \Tr_{(c)}\ln\left(-\neu{\hat{D}}
\neu{\hat{D}}\right) + \Tr_{(c,L)}\left(\trans\ln\HH\right) + 
\Tr_{(c,L)}\left(\longit\ln\longit\right).
\label{<30c>}
\end{equation}
Note that only the generalised transversal part of $\ln \HH$ contributes to
the effective action. Furthermore the last term in (\ref{<30c>}) can be 
interpreted as the entropy of the generalised longitudinal degrees of freedom.
In the perturbative calculation performed in section \ref{sec:3} we will
also see that the first term in (\ref{<30c>}) cancels against the 
FP determinant.

Eq.~(\ref{<22>}) represents the effective Abelian 
theory with the the non-Abelian effects contained in the action 
$S_{\rm int}$ (\ref{<25>}). This is as far as one can get exactly.
For the remaining integrations in (\ref{<25>}) over the auxiliary field 
$\chi$ (and the FP ghosts $\eta,\bar{\eta}$) we have to resort to some 
approximation. 
Previous applications of the path integral linearisation method suggest
that a semiclassical type of treatment in the auxiliary field $\chi$ is 
appropriate, since a good deal of the non-perturbative quantum 
effects have already been included by partly integrating out the 
original gauge fields.
We discuss the saddle point equation for $\chi$ in the next section.

\subsection{The Saddle-Point Equation}
\label{sec:2.3}
In order to perform the remaining integration in (\ref{<25>}) we are looking 
for the stationary points of the action {\sf S} (\ref{<26>}) with 
respect to the auxiliary fields $\chi,\eta$ and $\bar{\eta}$. The calculation 
of the corresponding functional derivatives is straightforward (though lengthy)
and we find the following equations of motion:
\begin{eqnarray}
\chi_{\mu\nu}(x) - i \hbar g^2 \epsilon^{\bar{a}\bar{b} 3}\, 
\DD^{\bar{a}\bar{b}}_{\mu\nu}(x,x) &=& 0 
\nonumber \\
\int\bar{\eta}^{\bar{a}}\cdot\Omega^{\bar{a}\bar{b}}[a,\chi,\eta,\bar{\eta}]
\cdot\eta^{\bar{b}} &=& 0
\label{<31>}
\end{eqnarray} 
Here, the kernel in the ghost eom.~is explicitly given by
\begin{equation}
\Omega^{\bar{a}\bar{b}}(x,y) := \left[\left[- 
\neu{D_\sigma} \neu{D_\sigma}\right]^{\bar{a}\bar{b}}(x) + \hbar g^2 
\DD^{\bar{a}\bar{b}}_{\mu\mu}(x,x) - \hbar g^2 \delta^{\bar{a}\bar{b}} 
\DD^{\bar{c}\bar{c}}_{\mu\mu}(x,x)\right]\,\delta(x,y) .
\label{<32>}
\end{equation}
The gap equations (\ref{<31>}) as well as their solutions 
depend of course on the Abelian gauge field $a_\mu$
and in lowest order saddle point approximation the photon interaction 
$S_{\rm int}$ from (\ref{<25>}) is given by the effective action
(\ref{<26>}) taken at the solution of (\ref{<31>}).  
Thus, the direct application of the saddle point method to the functional 
integral (\ref{<25>}) requires the solution of eq.~(\ref{<31>}) for 
{\em arbitrary} photon fields $a_\mu$.

To obtain the full YM partition function the remaining integral over the
Abelian field $a_\mu$ has to be carried out, too. 
In order to treat all fields on equal footing, one may wish to perform the 
$a_\mu$-integral in saddle point approximation as well. In this case we would
have to solve (\ref{<31>}) together with the $a_\mu$-equation (obtained by
varying $S_0 + S_{\rm int}$)
\begin{equation}
a_\mu + g^2\,[D_0]_{\mu\nu}\,\frac{\delta S_{\rm int}}{\delta a_\nu} = 0
\label{<33b>} .
\end{equation}
This system of equations is quite complicated to solve, in general, but 
one easily checks that it has the particular solution 
$a = \chi = \eta = \bar{\eta}= 0$.
The semiclassical expansion around this {\em trivial} saddle point corresponds 
to the standard perturbation theory as shown in the next chapter.

Some remarks on the trivial solution are in order. First of all, the 
ghost eom.~{\em always} (i.e.~independently of the value of $a$ and $\chi$)
allows for the solution $\eta = \bar{\eta} = 0$ and this is in fact the 
{\em only} one if the kernel $\Omega$ is invertible. The existence of 
a non-vanishing ghost condensate $\langle\bar{\eta}\eta\rangle \neq 0$ is not 
generally ruled out, but in the present paper we are mainly interested in the 
perturbative aspects of our formulation, and thus we confine ourselves to 
$\eta = \bar{\eta} = 0$. 

With the ghost fields set to zero, the remaining gap equation for $\chi$ 
always allows for the trivial solution $\chi = 0$. To see this, set 
$\chi = 0$ in the $a_\mu$-equation (\ref{<33b>}) and find $a_\mu = 0$.
This in turn implies that the charged gluon propagator becomes diagonal 
in colour space $\DD^{\bar{a}\bar{b}} \sim \delta^{\bar{a}\bar{b}}$, 
which also persists at any finite order of a semiclassical expansion for the
$a_\mu$-integral. Physically, it means that with this trivial 
saddle point, the charged gluon never changes its colour while propagating. 

Due to its inhomogeneity we expect the $\chi$-equation (\ref{<31>}) to allow 
for non-trivial solutions as well, describing non-perturbative
effects in the photon interaction. This is what happens in the so-called 
field strength approach \cite{R9}. However in 
the present work, we are mainly interested in perturbation theory, and
therefore we concentrate on the trivial solution $\chi \equiv 0$. Note that
the standard field strength approach leads to a strong coupling expansion
which does not allow for an immediate contact to perturbation theory.

In the rest of this paper, we will focus on the detailed description of the 
semiclassical type of expansion around the trivial saddle point. 
In particular, we will show how to recover the usual loop expansion in our 
formulation and we will calculate the $\beta$-function for the running 
coupling constant to one-loop order.

\section{The Loop Expansion}
\label{sec:3}
In this section we study the generating functional (\ref{<22>},\ref{<25>})
in a semiclassical expansion around the trivial saddle point discussed
in the last section. 

\subsection{The Effective Action in a Semiclassical Expansion}
\label{sec:3.1}
Our starting point is the generating functional for the full YM theory,
as obtained by inserting the effective photon action (\ref{<25>}) in 
eq.~(\ref{<22>}):
\[
Z[j,\ldots] = \int {\cal D}(a,\chi,\eta,\bar{\eta})\exp\left\{-\frac{1}{\hbar}
\left(S_0+{\sf S}\right) + \mbox{cur.}\right\}.
\] 
Here, "cur." stands for the current terms in the exponent of (\ref{<25>})
plus the $\neu{j_\mu}\cdot a_\mu$ coupling from (\ref{<22>}).   

The following calculation is somewhat cumbersome due to the number of
integration variables in $Z$. We will use the superfield notation 
\cite{git} and collect the integration variables in the symbol
$\Theta = \{a,\eta,\bar{\eta},\chi\}$ with the corresponding 
currents denoted by $\theta = \{\neu{j},\bar{\xi},\xi,\vartheta\}$\footnote{A 
source $\vartheta_{\mu\nu}$ for the neutral field $\chi_{\mu\nu}$ has been 
introduced for notational reasons.}. The generating
functional depends on two additional sources $\lambda=\{\ch{j_\mu},\varphi\}$
referring to the fields $\Lambda = \{\ch{A_\mu},\phi\}$ that have been 
integrated out in our approach. With this compact superfield notation, we can
re-express the generating functional as 
\begin{equation}
Z[\theta,\lambda] = \int {\cal D}\Theta\exp\left\{-\frac{1}{\hbar g^2}
\left({\cal F} + {\cal F}_{\rm cur}\right)\right\} ,
\label{<200>}
\end{equation}
where ${\cal F}[\Theta] = g^2 (S_0 + {\sf S})$ is the part of the action 
independent of the currents. The explicit forms for ${\cal F}$ and 
${\cal F}_{\rm cur}$ can be found in appendix \ref{app:2}.

Our goal is to perform the superfield integration in eq.~(\ref{<200>}) in 
a semiclassical expansion around the {\em trivial} saddle point $\bar{\Theta}$
of the exponent. Since we include the current terms in the action the 
saddle point itself will be a functional of the currents, $\bar{\Theta} = 
\bar{\Theta}[\theta,\lambda]$. The explicit form of this dependence is not
needed, however, for the calculation below. All we require is
that at zero currents, the saddle point is also zero: 
$\bar{\Theta}[\theta = \lambda = 0] = 0$. The existence of this {\em trivial}
saddle point at zero currents was shown in the last section.

Let us now expand the exponent ${\cal F} + {\cal F}_{\rm cur}$ of (\ref{<200>})
to second order around the saddle point $\bar{\Theta}$ and perform the 
Gaussian superfield integration. Taking the logarithm, 
$W = - \hbar g^2 \ln Z$, we obtain the generating functional for connected 
Green's functions:
\begin{equation}
W[\theta,\lambda] = {\cal F}[\bar{\Theta}] + {\cal F}_{\rm cur}[\bar{\Theta},
\theta,\lambda] + \frac{\hbar g^2}{2}\Tr\ln\left.\frac{\delta^2 ({\cal F}+
{\cal F}_{\rm cur})}{\delta\Theta^2}\right|_{\Theta = \bar{\Theta}} + 
{\cal O}(\hbar^2) .
\label{<202>}
\end{equation}
The supertrace term on the rhs.~is the next to leading correction in our 
semiclassical expansion. It contains terms of order ${\cal O}(\hbar)$ and 
${\cal O}(\hbar^2)$, i.e.~the expansion around the trivial saddle point is
{\em not} directly organised in powers of $\hbar$. What we claim, however, is
that to a given order in the saddle point approximation (SPA), we find {\em all}
the standard loop corrections to this order, plus some higher contributions in 
$\hbar$. Thus, given the SPA to a fixed order, we can rearrange it in powers of
$\hbar$ such that we recover the loop series.   

In the formula (\ref{<202>}), the explicit dependence of the saddle point 
$\bar{\Theta}$ on the currents is needed. This can be avoided by performing a
Legendre transformation to obtain the effective action $\Gamma$ as generating 
functional for 1PI Green's functions. 
In principle, this transformation is straightforward: First we define classical
superfields $\vek{\Theta},\vek{\Lambda}$ as functionals of the currents:
\begin{equation}
\vek{\Theta}[\theta,\lambda] = - \frac{\delta W[\theta,\lambda]}
{\delta\theta} \quad , \quad \vek{\Lambda}[\theta,\lambda] = - \frac{\delta
W[\theta,\lambda]}{\delta\lambda}.
\label{<203>}
\end{equation}  
Remember that these superfields stand for the component classical fields
$\vek{\Theta} = \{{\sf a}_\mu,\vek{\eta},\bar{\vek{\eta}},\vek{\chi}_{\mu\nu}\}$
and $\vek{\Lambda} = \{{\sf A}_\mu,\vek{\phi}\}$. With this definition, 
the effective action can be written as
\begin{equation}
 g^2 \Gamma[\vek{\Theta},\vek{\Lambda}] =W[\theta[\vek{\Theta},\vek{\Lambda}],
 \lambda[\vek{\Theta},\vek{\Lambda}]] + \int \vek{\Theta} \theta[\vek{\Theta},
 \vek{\Lambda}] + \int \vek{\Lambda} \lambda[\vek{\Theta},\vek{\Lambda}].
\label{<204>}
\end{equation}
Following the remarks above, we will rearrange the expression (\ref{<202>})
for $W$ in powers of $\hbar$ and calculate the effective action in a 
$\hbar$-expansion, $\Gamma = \Gamma_0 + \hbar \Gamma_1 + \cdots$. We start by
inserting (\ref{<202>}) in the definition of the classical superfields 
(\ref{<203>}) and expand in $\hbar$:
\begin{equation}
\vek{\Theta}[\theta,\lambda] = \bar{\Theta}[\theta,\lambda] + {\cal O}(\hbar)
\quad , \quad \vek{\Lambda}[\theta,\lambda] = - \left.\frac{\delta ({\cal F}+
{\cal F}_{\rm cur})}{\delta\lambda}\right|_{\Theta = \bar{\Theta}} + 
{\cal O}(\hbar) .
\label{<205>}
\end{equation}
The first equation is what one usually expects: The classical field coincides
with the saddle point value to lowest order. This fact can be used to 
simplify the calculation of the effective action. When inserting our result 
(\ref{<202>}) for $W$ in the definition (\ref{<204>}) of $\Gamma$ and expanding
in powers of $\hbar$, we can simply replace $\bar{\Theta}\rightarrow
\vek{\Theta}$ to one-loop order and the saddle point $\bar{\Theta}$ drops out 
of the calculation. 

The final step is now to eliminate in $\Gamma$ the currents $\theta$ and 
$\lambda$ in favour of the classical fields by inverting (\ref{<203>}).
Recall that the fields $\Lambda=\{\ch{A_\mu},\phi\}$ have been integrated out, 
but the generating functional still contains the corresponding currents 
$\lambda=\{\ch{j_\mu},\varphi\}$. Therefore, the second
eq.~(\ref{<203>}) defines classical fields $\vek{\Lambda} = \{{\sf A}_\mu,
\vek{\phi}\}$ which should not be confused with the original integration
variables. The current for the auxiliary field $\chi$ has been introduced
for notational reasons only and the corresponding classical field $\vek{\chi}$ 
is therefore not an independent quantity. By the path integral linearisation
(\ref{<20>}), $\chi$ is the field conjugate to the commutator
term $[\ch{A_\mu},\ch{A_\nu}]$. In terms of the classical field ${\sf A}_\mu$,
we therefore expect to lowest order:
\begin{equation}
\vek{\chi} =  i \epsilon^{\bar{a}\bar{b}3}{\sf A}_\mu^{\bar{a}}
{\sf A}_\nu^{\bar{b}} + {\cal O}(\hbar) .
\label{<207>}
\end{equation}
The explicit derivation of (\ref{<207>}) can be found  in appendix \ref{app:2},
together with the elimination of the currents in $\Gamma$.
As our final result we find the following structure: 
\medskip

The lowest oder part of the effective action coincides precisely with the
initial BRST-action (\ref{<14>}), i.e.
\begin{equation}
\Gamma_0[{\sf a}_\mu, {\sf A}_\mu,\bar{\vek{\eta}},\vek{\eta},\vek{\phi}]
= S_{\rm eff}[{\sf a}_\mu, {\sf A}_\mu,\bar{\vek{\eta}},\vek{\eta},\vek{\phi}]
\qquad \mbox{see (\ref{<14>})}\,\,.
\end{equation}
Since we have completely reformulated the non-Abelian part of the theory,
this is a non-trivial test for the calculation. Our main interest, however,
lies in the one-loop corrections, which are given by
\begin{equation}
\Gamma_1 = \frac{1}{2} \Tr \ln \HH + \frac{1}{2} \Tr \ln \LL
+ \frac{1}{2} \Tr\ln \left.\frac{\delta^2 {\sf F}}{\delta \Theta^2}
\right|_{\vek{\Theta}} .
\label{<206>}
\end{equation}
The arguments of the operators in this formula are the classical fields 
(cf.~(\ref{<207>})). 

The operator ${\sf F}$ in the supertrace on the rhs.~of (\ref{<206>}) is 
quite complicated, see appendix \ref{app:2}. This is mainly due to the fact 
that it explicitly contains the currents $\lambda = \{\ch{j_\mu},\varphi\}$
as functional of the classical fields, leading to rather messy expressions.
We refrain from presenting them in any detail here.

At this point, we can explore the fact that we are expanding around the
{\em trivial} saddle point: At zero currents, the classical fields are 
zero and vice versa. This means that in the effective action we can simply set
the complicated currents $\lambda = \{\ch{j_\mu},\varphi\} = 0$ if the 
corresponding classical fields ${\sf A}_\mu,\vek{\phi}$ vanish. Thus, we 
restrict ourselves to 1PI diagrams with
no external charged gluons and multiplier lines. For the purpose of 
calculating the $\beta$-function, this is sufficient if we choose to define the 
coupling constant through the ghost-photon-vertex. By BRST-invariance, i.e.~the
Slavnov-Taylor identities, any other definition must give the same answer. 
In this special case, we can actually expand the supertrace in 
(\ref{<206>}) in its component blocks with the result\footnote{From the 
$(\bar{\eta},\eta)$ block, we obtain the covariant Laplacian, $-\Tr\ln
\left(-\neu{D}\neu{D}\right)$, where the minus sign originates from the 
fermion nature of the ghost loop. As mentioned at the end of section 
\ref{sec:3.2}, this "perturbative" FP determinant cancels against the
corresponding term in (\ref{<30c>}).}
\begin{eqnarray}
\Gamma_1[{\sf a}_\mu,\bar{\vek{\eta}},\vek{\eta}, {\sf A}_\mu = 0, \vek{\phi}
= 0] &=& \frac{1}{2} \Tr \ln \HH + \frac{1}{2} \Tr \ln \LL - 
\Tr\ln\left[-\neu{D_\sigma}\neu{D_\sigma}\right] +
\nonumber \\
&& {}\hspace{-2cm} + \frac{1}{2} \Tr \ln \left[\left(D_0^{-1}\right)_{\mu\nu} 
+ 2\,\delta_{\mu\nu}\bar{\vek{\eta}}^{\bar{a}} \vek{\eta}^{\bar{a}} - 
2\,\Delta_{\mu\nu} [\bar{\vek{\eta}},\vek{\eta},{\sf a}]
\right]
\label{<208>}
\end{eqnarray}
where $\Delta_{\mu\nu}$ is a quadratic form in the ghost fields, which is 
given in appendix \ref{app:2}.

Let us finally consider the case that we only have external photon lines,
i.e.~examine the effective photon theory. In this case, we only retain
the photon field in the effective action and obtain the simple one-loop result
\begin{eqnarray}
\Gamma_0[{\sf a}_\mu] &=& \frac{1}{2 g^2} \int {\sf a}_\mu
\left[D_0^{-1}\right]_{\mu\nu} {\sf a}_\nu \,d^4 x
\nonumber \\
\Gamma_1[{\sf a}_\mu] &=& \frac{1}{2} \Tr \ln \HH + \frac{1}{2} \Tr \ln \LL
- \Tr \ln  \left[-\neu{D_\sigma}\neu{D_\sigma}\right] 
\label{<210>}
\end{eqnarray}
with all operators taken at $\chi = \eta = \bar{\eta} = 0$.

\subsection{The {\sf MAG} symmetry}
\label{sec:3.2}
The {\sf MAG} looks very similar to a background gauge with the neutral
photon playing the role of the background field. This suggests that
the $\beta$-function for the running coupling constant 
only depends on the photon propagator, if one chooses to define the 
coupling constant through a vertex involving $a_\mu$. In this section,
we will prove this statement explicitly for the 
$\bar{\eta}\eta a_\mu$-vertex,\footnote{We have checked this statement
for the $\ch{A}\ch{A}a$ triple gluon vertex as well. As explained above
the explicit calculations are rather cumbersome in this case and were 
in fact performed with the help of a symbolic algebra program. In order to 
present the formulae as simple as possible, we take the ghost-photon vertex
to define the coupling constant.}
i.e.~we show that the one-loop corrections to this vertex are exactly
cancelled by the corrections to the ghost propagator, so that the 
coupling $g$ defined by this vertex is exclusively renormalised by the neutral 
vacuum polarisation:\footnote{Here
and in the following the wavy and the dashed lines refer to gluons (neutral 
or charged as indicated) and ghosts, respectively.}
\[
\begin{minipage}{10.5cm}
\epsfig{file = 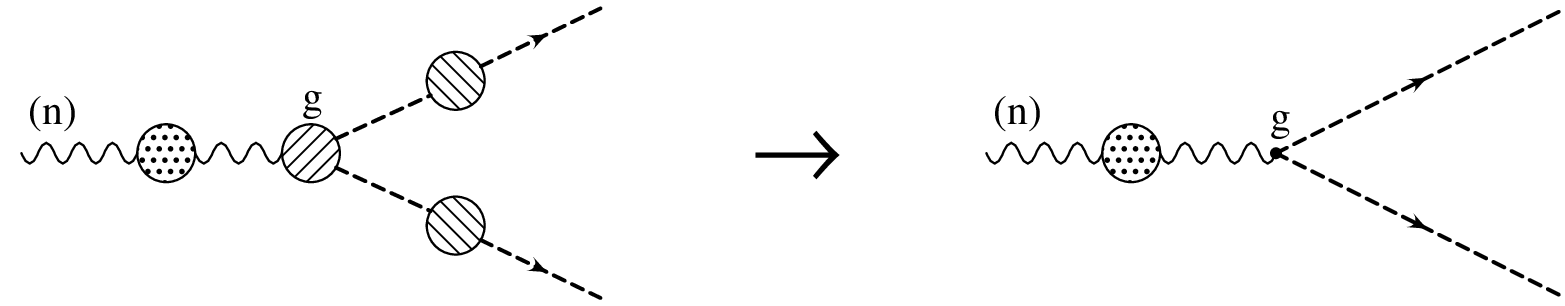,width = 10cm}
\end{minipage} 
\]
To this end, we keep only the ghost and photon fields in the effective 
action, i.e.~we start from (\ref{<208>}).
 
For the computation, we proceed as follows: The $n$-point functions are
obtained as functional derivatives of the effective action calculated in 
the last section. 
For a typical term $\Gamma_1 \simeq \Tr \ln \Omega$ ($\Omega$ some operator) 
this gives the trace of a sequence of {\em propagators} $\Omega^{-1}$ and 
{\em vertices} (derivatives of $\Omega$), both taken at zero fields.
Introducing momenta by a Fourier representation, all the spacetime integrations
in the trace can be performed. In this way we recover the usual Feynman
diagrams in momentum space. Some cautionary remarks on the remaining loop
integration over the internal momentum are in order:

In the calculation below, we will frequently encounter loop integrations which
are not only logarithmically, but also linearly and quadratically divergent
(by power counting). The counterterms for this latter divergences are
ambiguous, i.e.~they depend on the regularisation method, and, even worse, 
they correspond to terms (e.g.~photon mass terms) that are not present in 
the initial Lagrangian. The reason for this problem is clear: The residual 
$U(1)$ gauge invariance (\ref{<24a>}) is "artificially" broken by the 
regularisation scheme. As a consequence, the Green's functions are not of the 
form required by the Ward identities. 

There are two possibilities to cure this situation: In a regularisation 
independent scheme like BPHZ, one subtracts off a sufficient number of 
leading terms in the Taylor expansion around zero momentum. This removes the 
dangerous ambiguities. A more obvious method is to use a gauge invariant
regularisation scheme from the very beginning. At least within the class of 
such regularisations, the results should be unambiguous and the dangerous 
divergences should simply drop out.   

We have performed our calculations in two independent, gauge invariant
regularisation schemes: In the main text, we will use dimensional 
regularisation to $d = 4-\epsilon$ spacetime dimensions. There is one problem
with this scheme, namely the fact that quadratic divergences are automatically
set to zero. Thus, this scheme does not allow for a check whether the 
dangerous terms really {\em cancel}. In order to parametrise the ambiguities
(rather then setting them to zero from the outset), we have redone all the
calculations using {\em proper time regularisation}, which preserves 
gauge invariance. This also gives a check on the scheme independence mentioned
above.    

Finally we should also remark that formally correct operations with divergent 
integrals can lead to ambiguous results. While we can always come to finite
expressions by a suitable cutoff procedure, some arithmetic operations may
be in conflict with the regularisation. In all
of our calculations, we have only performed such manipulations that are allowed
by the respective regularisation scheme. In practice, the main feature of 
a gauge invariant regularisation is the possibility to shift the loop momentum.

Some more details on the regularisation procedure as well as some results
of the proper time calculation can be found in appendix \ref{app:3}.

Let us now present the results of the explicit calculation:\\[3mm]
{\bf 1. Ghost vacuum polarisation:} The loop corrections to the ghost
propagator are defined by\footnote{Here and in the following, all 
derivatives with respect to Grassmann fields $\bar{\eta}$ or $\eta$ are taken 
to be {\em left} or {\em right} derivatives, as indicated.}
\begin{equation}
\left\langle\bar{\eta}^{\bar{a}}(x)\eta^{\bar{b}}(y)\right\rangle = 
\hbar \left[\frac{\delta^2 \Gamma}{\delta_r\eta^{\bar{b}}(y)\delta_l
\bar{\eta}^{\bar{a}}(x)}\right]^{-1} = \hbar g^2 \left[ \MM_0^{-1} - 
\MM_0^{-1}\cdot\Pi^{(\rm gh)}\cdot \MM_0^{-1} + \cdots\right],
\end{equation}
where the free propagator is 
$\MM_0^{\bar{a}\bar{b}}(p) = \delta^{\bar{a}\bar{b}}/p^2$ 
(see eq.~(\ref{<13>})) and   
\begin{equation}
\Pi^{(\rm gh)}_{\bar{a}\bar{b}}(x,y) = \left.\hbar g^2 \frac{\delta \Gamma_1}
{\delta_r\eta^{\bar{b}}(y)\delta_l\bar{\eta}^{\bar{a}}(x)}\right|_0
\end{equation}
is the one-loop correction.

There are three contributions from (\ref{<208>}) that are proportional to 
$\int \frac{d^d k}{k^2}$ and vanish in dimensional regularisation, 
namely the $\HH$ and $\LL$ contribution and the piece coming from the 
second term in the last trace (\ref{<208>}). 
In addition there is a contribution from the complicated $\Delta$-term
in the last trace (\ref{<208>}). In momentum space, it reads
\begin{eqnarray}
\Pi^{({\rm gh})}_{\bar{a}\bar{b}}(p) &=& - \hbar g^2 \mu^\epsilon 
\,\delta^{\bar{a}\bar{b}}\int \frac{d^d k}{(2\pi)^d} \left[D_0\right]_{\nu\mu}
(k)\,\, \frac{(2p+k)_\mu (2p+k)_\nu}{(k+p)^2}
\nonumber \\
&=& \hbar g^2\,\delta^{\bar{a}\bar{b}}\,\frac{\alpha-3}{8 \pi^2 \epsilon}
\,p^2 + {\rm finite}\,\,.
\label{A1}
\end{eqnarray} 
where the arbitrary scale $\mu$ must be introduced to keep the renormalised 
coupling constant dimensionless in $d\neq 4$.
Note that this non-vanishing contribution can be identified with the diagram
\[
\begin{minipage}{3.2cm}
\epsfig{file = 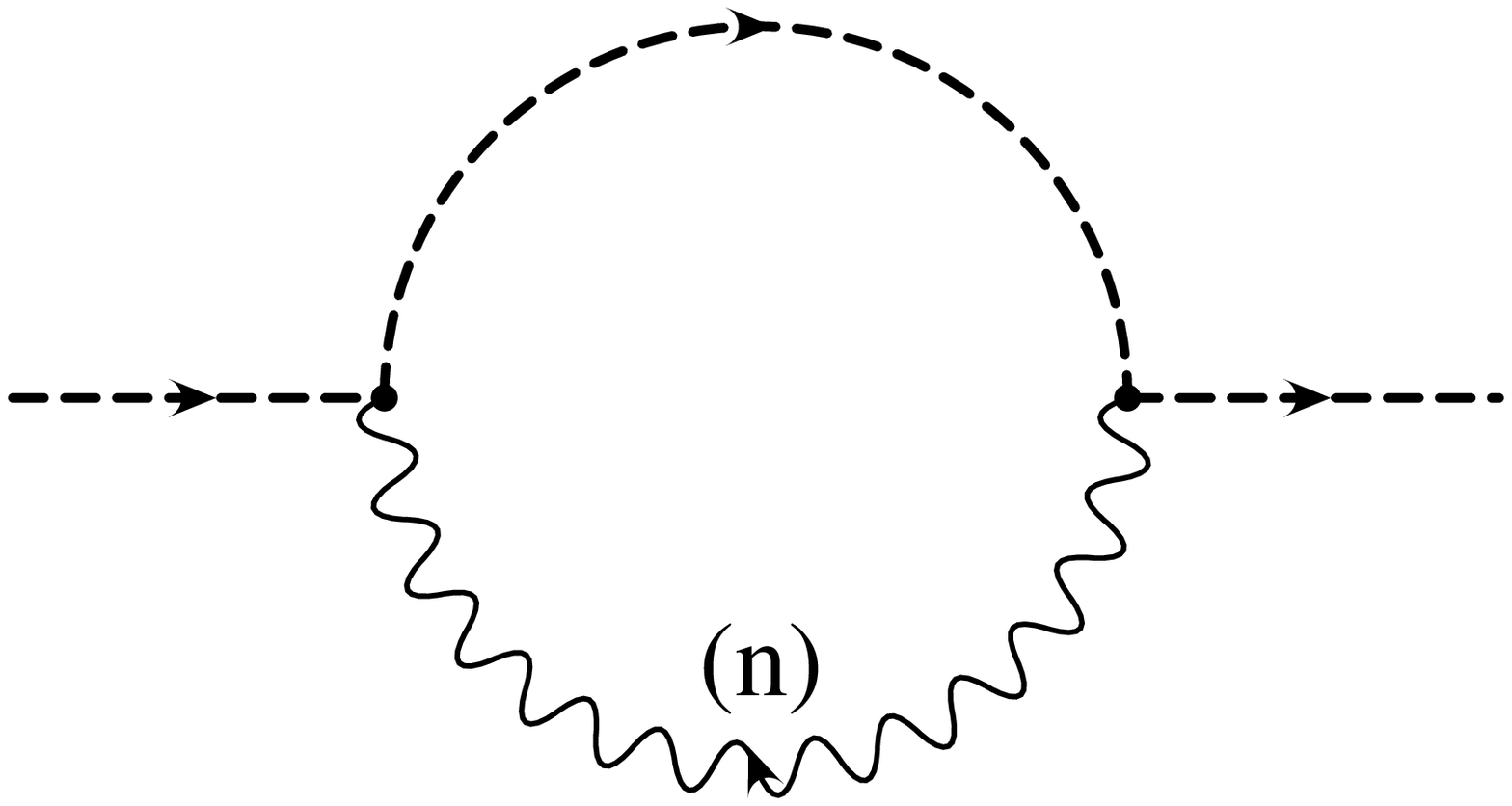,width = 3.0cm}
\end{minipage}. 
\] 
There is a quadratic divergence in (\ref{A1}) which cancels exactly against
the three contributions $\sim\int \frac{d^d k}{k^2}$ discussed above, if
we parametrise it in a gauge invariant way. See the proper time calculation 
given in appendix \ref{app:3} for an example of such a cancellation.

The finite part in (\ref{A1}) depends on the arbitrary scale $\mu$, whereas 
the divergent part does not. As a consequence, the corresponding counterterm 
in the MS-scheme,
\begin{equation}
\delta \Gamma = \frac{1}{g^2 \mu^\epsilon}\int \bar{\eta}^{\bar{a}} 
\cdot\left[\left. -\Pi^{\bar{a}\bar{b}}_{\rm (gh)}\right|_{\rm div}
\right] \cdot\eta^{\bar{b}} 
= \frac{1}{g^2 \mu^\epsilon}\int \bar{\eta}^{\bar{a}} \left(-
\dalembert\right) \eta^{\bar{a}}\cdot \left(\hbar g^2 \frac{3-\alpha}
{8 \pi^2\epsilon}\right)
\label{AA}
\end{equation}
depends on the scale $\mu$ only by the overall prefactor. 
Since (\ref{AA}) is proportional to the free ghost term in the initial
BRST action (in $d$ dimensions), we can renormalise {\em multiplicatively}.  
Our definitions of renormalisation constants are
\begin{eqnarray}
g_0\,\eta_0 &=& Z^{1/2}_{\rm gh}\cdot g \mu^{\epsilon/2}\,\eta \qquad ; \qquad
g_0\,\bar{\eta}_0 = Z^{1/2}_{\rm gh}\cdot g \mu^{\epsilon/2}\,\bar{\eta} 
\nonumber\\
g_0\, a_\mu^{(0)} &=& Z_n^{1/2}\cdot g \mu^{\epsilon/2}\, a_\mu \label{Z} \\
g_0 &=& g \mu^{\epsilon/2} \cdot Z_g Z_n^{-1/2}\, Z_{\rm gh}^{-1}
\nonumber
\end{eqnarray}
where the bare quantities are denoted by an index '0'. In perturbation theory
one usually rescales the fields by a factor of $g$ (as compared to our
conventions) in order to get rid of the prefactor $1/(g^2\mu^\epsilon)$ in the 
action. We have inserted extra factors of $g$ and $g_0$ in the definition of 
the wave-function renormalisation such that the choice (\ref{Z}) gives the same
$Z$'s as with the perturbative convention.  From (\ref{AA}) and (\ref{Z}), we 
can read off $Z_{\rm gh}$:
\begin{equation}
Z_{\rm gh} = 1 - \hbar g^2\,\frac{\alpha-3}{8 \pi^2 \epsilon} + \cdots
\label{<209>}
\end{equation}
\\[3mm]
{\bf 2. Ghost-photon vertex:} As usual, we define the vertex through the
amputated three-point function 
\begin{equation}
- g^2 \left.\frac{\delta^3 \Gamma}{\delta_r\eta^{\bar{b}}(z) 
\delta_l\bar{\eta}^{\bar{a}}(y)\delta a_\mu(x)}\right|_0 \equiv 
i \epsilon^{\bar{a}\bar{b} 3} \int \frac{d^4(p,q)}{(2\pi)^8} 
e^{-i p(x-y) - i q(x-z)}\,G_\mu(p,q).
\label{1000}
\end{equation}
The bare vertex from the initial BRST-action is
\begin{equation}
G_\mu^{(0)}(p,q) = q_\mu - p_\mu .
\label{<250>}
\end{equation}
By power counting, the one-loop corrections to this vertex are linearly 
divergent. Note however, that the results are unambiguous in any gauge 
invariant regularisation scheme, i.e.~if we can shift the loop
momentum in the regularised integral.  

As an example, consider the contribution from the first trace in (\ref{<208>}).
This correction to $G_\mu(p,q)$ can be identified with the following diagram:
\begin{equation}
\begin{minipage}{3.5cm}
\epsfig{file = 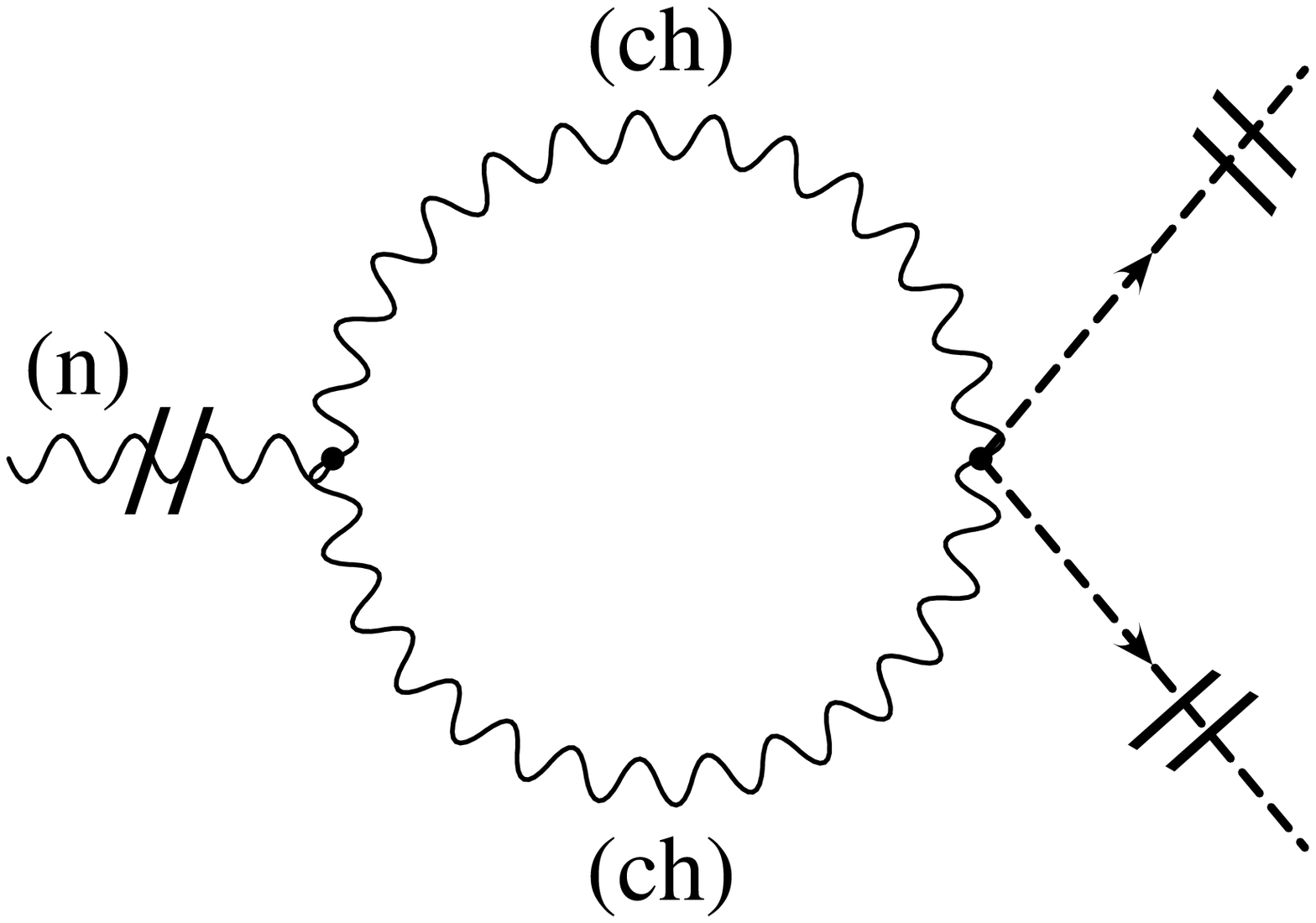,width = 3cm}
\end{minipage} 
= - \hbar g^2 d \int\frac{d^d k}{(2\pi)^d} \frac{(2k+p+q)_\mu}
{k^2 (k+p+q)^2} = 0.
\label{A4}
\end{equation}
Clearly, this will vanish in {\em any} regularisation scheme that allows for 
the shift $k\mapsto p+q-k$ in the loop momentum. Note that the 
diagram (\ref{A4}) manifestly is a function of $p+q$ and thus {\em not}
proportional to the bare vertex (\ref{<250>}) as requested by the Ward 
identities. It is therefore clear that this diagram has to vanish in any gauge 
invariant regularisation scheme.

The same analysis also applies to the contribution from the second trace in 
(\ref{<208>}). The calculations are a bit more involved, but the resulting
loop integral vanishes identically in any gauge invariant scheme, just as
in (\ref{A4}).
 
Let us now look at the non-vanishing loop corrections to the vertex. They
come from the last trace in the effective action (\ref{<208>}) and split into
two parts which can be identified with the following diagrams: 
\begin{equation}
\begin{minipage}{8.5cm}
\epsfig{file = 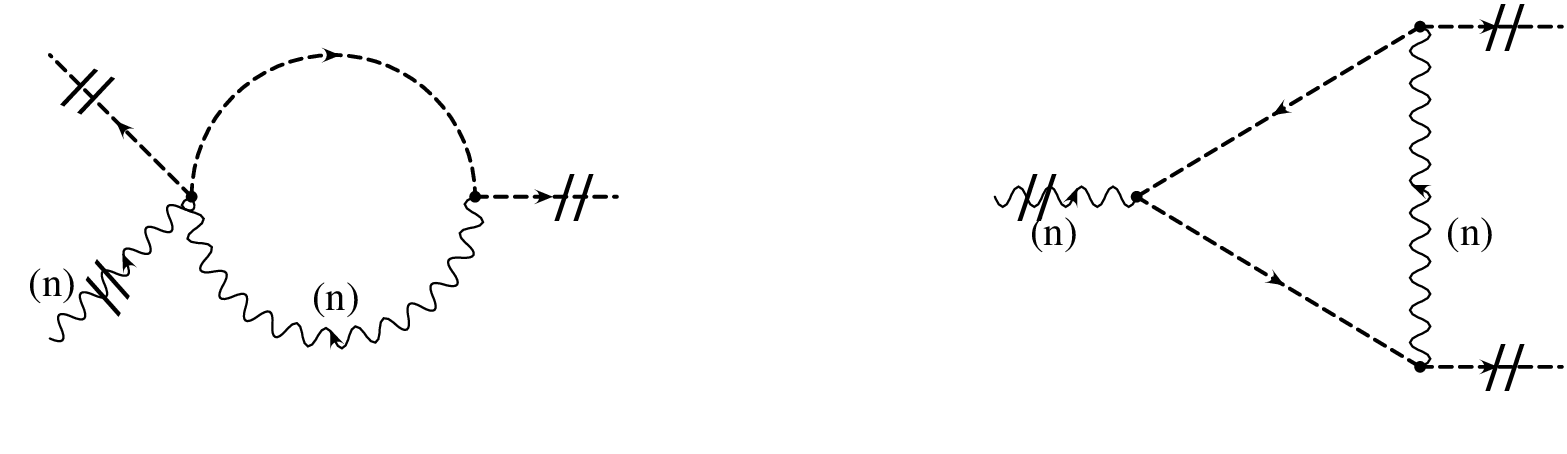,width = 8cm}
\end{minipage}. 
\end{equation}
Each of these diagrams is proportional to the bare vertex (\ref{<250>}). We 
find
\[
2 \hbar g^2 \mu^\epsilon \int\frac{d^d k}{(2\pi)^d} [D_0]_{\mu\nu}(k)\left[
\frac{(2p-k)_\nu}{(k-p)^2} - \frac{(2q+k)_\nu}{(k+q)^2}\right] = (q-p)_\mu\cdot\hbar g^2
\frac{-3}{8 \pi^2\epsilon} + \mbox{finite}
\]
for the left diagram. The triangle diagram vanishes in Lorentz gauge for the
internal photon line:
\[
- \hbar g^2 \mu^\epsilon\int\frac{d^d k}{(2\pi)^d} [D_0]_{\sigma\tau}(k)
\frac{(2p-k)_\sigma (2q+k)_\tau (2k+q-p)_\mu}{(k-p)^2 (k+q)^2}
= (q-p)_\mu\cdot\hbar g^2 \frac{\alpha}{8 \pi^2\epsilon} + \cdots .
\]
Note that in the proper time formalism we would find the same results with 
the substitution $1/\epsilon \leftrightarrow 1/2\ln(\Lambda^2/p^2)$.

To construct a counterterm from these diagrams in the MS scheme, we add the 
negative divergences to the initial vertex:
\begin{equation}
G_\mu(p,q) \equiv G_\mu^{(0)}(p,q) \cdot Z_g \qquad ; \qquad 
Z_g = 1 + \hbar g^2\,\frac{3-\alpha}{8 \pi^2 \epsilon} + \cdots .
\end{equation}
The counterterm is then given by Fourier transforming and multiplying with 
external fields to undo the differentiation in (\ref{1000}). This brings in 
additional $Z$-factors and yields the relation (\ref{Z}) between the bare and 
renormalised coupling constants. 

The cancellation condition $Z_g = Z_{\rm gh}$ means that the coupling 
constant defined through the ghost-photon vertex is only renormalised due to 
the photon vacuum polarisation. The latter is therefore sufficient to 
determine the $\beta$-function in the ${\sf MAG}$.

\subsection{The Photon Vacuum Polarisation}
\label{sec:3.3}
We have seen in the last section that the running coupling constant
in the {\sf MAG} is determined by the renormalisation of the Abelian gluon 
propagator alone. 
To calculate the correction to the photon propagator, we only retain the 
photon field in the one-loop effective action, i.e.~we start from 
(\ref{<210>}). 

The vacuum polarisation is connected with the full two-point function
by
\begin{eqnarray*}
\left<a_\mu(x) a_\nu(y)\right> = \hbar \left[\left.\frac{\delta^2 \Gamma}
{\delta {\sf a}_\mu(x) \delta{\sf a}_\nu (y)}\right|_0\right]^{-1} 
&=& \hbar \left[ \frac{1}{g^2} \left(D_0^{-1}\right)_{\mu\nu} + 
\frac{1}{g^2} \neu{\Pi_{\mu\nu}}\right]^{-1} \\
&=& 
\hbar g^2 \left[ D_0 - D_0 \neu{\Pi} D_0 + \cdots\right]_{\mu\nu},
\end{eqnarray*}
which in turn leads to 
\begin{equation}
\neu{\Pi_{\mu\nu}}(x,y) \equiv 
\hbar g^2 \left. \frac{\delta^2 \Gamma_1}
{\delta {\sf a}_\mu(x) {\sf a}_\nu(y)}\right|_0.
\label{<101>}
\end{equation} 
Note that $\neu{\Pi}_{\mu\nu}$ is transversal in any gauge invariant 
regularisation scheme, as a general consequence of the Ward identities 
from the residual $U(1)$ symmetry (\ref{<24a>}). 

Let us now consider the contribution from the three determinants in the
effective photon action (\ref{<210>}).

The determinant of the covariant Laplacian is well known from e.g.~heat kernel 
expansions. In momentum space, we find
\begin{eqnarray}
\neu{\Pi_{\mu\nu}}(p) &\stackrel{\rm lapl.}{=}& - \hbar g^2 \mu^\epsilon
\kappa\,\left[2 \delta_{\mu\nu} 
\int \frac{d^d k}{(2 \pi)^d} \frac{1}{k^2} - \int \frac{d^d k}
{(2 \pi)^d} \frac{(2 k + p)_\mu (2 k + p)_\nu}{k^2 (k+p)^2} \right]
\nonumber \\
{}& = & - \hbar g^2 \frac{\kappa}{24 \pi^2 \epsilon}\left[\delta_{\mu\nu} p^2 - 
p_\mu p_\nu\right] + \mbox{finite}.
\label{<104>}
\end{eqnarray}
The colour factor 
$
\kappa = \delta^{\bar{c}\bar{c}} = \epsilon^{\bar{a}\bar{b} 3}
\epsilon^{\bar{a}\bar{b} 3} = 2
$ 
has been introduced for later comparison with the $SU(N)$ 
result.\footnote{From $\epsilon^{a b c} \epsilon^{a b d} = 
\delta^{c d} \kappa$, the factor $\kappa$ is simply the quadratic Casimir
of the adjoint rep.~of $SU(2)$.}
The formula (\ref{<104>}) corresponds to the two diagrams
\[
\begin{minipage}{6.8cm}
\epsfig{file = 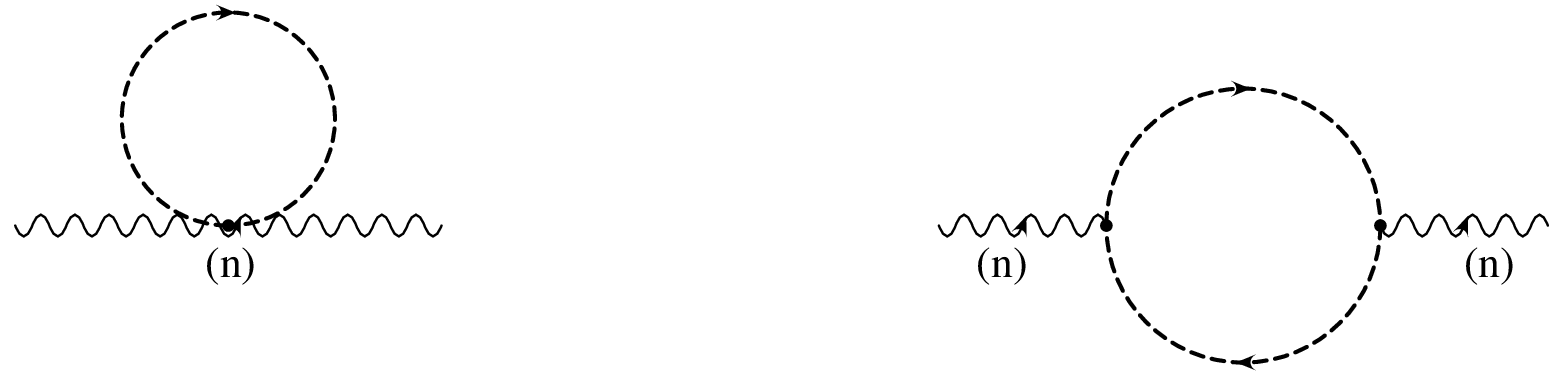,width = 6.5cm}
\end{minipage} 
\]
where the first one originates from the non-linearity of the {\sf MAG}.
If we regularise these diagrams using the proper time method (cf.~appendix 
\ref{app:3}), we find an exact cancellation of the inherent quadratic 
divergences between the two diagrams.  
The remaining logarithmic divergence leads to a transversal
result, as expected. 

The $\HH$-determinant in (\ref{<210>}) describes the (unprojected)
propagation of charged gluons, as expressed in the diagrams 
\[
\begin{minipage}{6.8cm}
\epsfig{file = 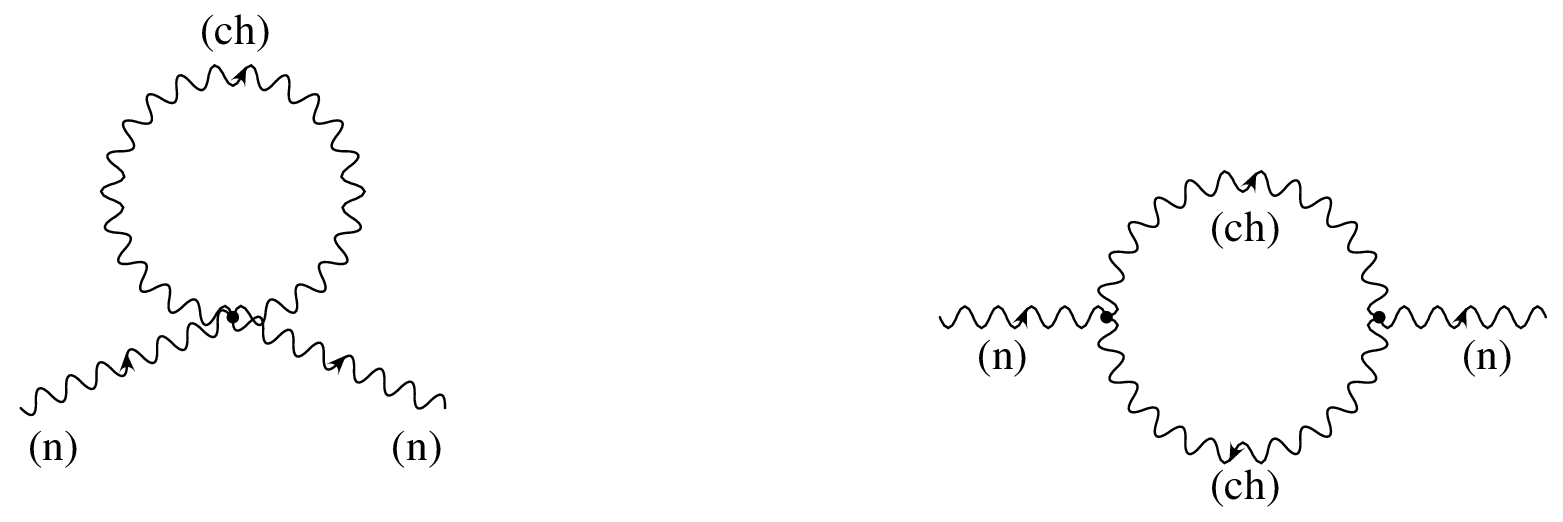,width = 6.5cm}
\end{minipage}. 
\]
The analytical expression is
\begin{eqnarray}
\neu{\Pi_{\mu\nu}}(p) &\stackrel{\sf H}{=}& \hbar g^2 \mu^\epsilon\kappa\,
\delta_{\mu\nu}\int \frac{d^d k}{(2\pi)^d} \frac{d}{k^2} -
\nonumber \\
{}&& - \hbar g^2 \mu^\epsilon \frac{\kappa}{2}
\int \frac{d^d k}{(2\pi)^d} \frac{8\, \delta_{\mu\nu} p^2 + 4 d k_\mu k_\nu
+ 2 d (k_\mu p_\nu + k_\nu p_\mu) + (d-8) p_\mu p_\nu}{k^2 (k+p)^2}
\nonumber \\
{} &=& - \hbar g^2 \kappa \left[ \frac{5}{12 \pi^2 \epsilon} \left( p^2 
\delta_{\mu\nu} - p_\mu p_\nu \right) + \cdots \right] . 
\label{<105>}
\end{eqnarray}
As before, the proper time calculation of appendix \ref{app:3} confirms that
the quadratic divergences cancel between the two diagrams, leaving us with 
the  unambiguous (and transversal) result (\ref{<105>}).

Finally, let us consider the $\LL$-determinant in the effective action
(\ref{<210>}), which describes the effect of the multiplier field, i.e.~the 
projection onto {\sf MAG}. The calculation turns out to be very tedious, but 
for completeness, let us quote the final result: 
\begin{eqnarray}
\neu{\Pi_{\mu\nu}}(p) &\stackrel{\sf L}{=}& \frac{\kappa}{2} \hbar g^2
\mu^\epsilon \int \frac{d^d k}{(2\pi)^d} \left[\frac{{\cal R}_{\mu\nu}(p,k)}
{k^2 (k+p)^2}+\frac{{\cal S}_{\mu\nu}(p,k)}{k^4 (k+p)^2}
+\frac{{\cal T}_{\mu\nu}(p,k)}{k^4 (k+p)^4} \right] 
\nonumber \\
{\cal R}_{\mu\nu}(p,k) &=& -2 (k_\mu p_\nu + k_\nu p_\mu) + 2 p_\mu p_\nu + 8 
\delta_{\mu\nu} (k\cdot p)
\nonumber \\
{\cal S}_{\mu\nu}(p,k) &=& 8\left[p^2 k_\mu k_\nu - (k\cdot p)(k_\mu p_\nu +
k_\nu p_\mu) + \delta_{\mu\nu} (k\cdot p)^2\right]
\nonumber \\
{\cal T}_{\mu\nu}(p,k) &=& - p^4 k_\mu k_\nu + p^2 (k\cdot p)(k_\mu p_\nu +
k_\nu p_\mu) - p_\mu p_\nu (k\cdot p)^2 .
\label{<106>}
\end{eqnarray}
All the divergences in this expression cancel, giving a finite 
result.\footnote{It might very well be zero in any gauge invariant 
regularisation, but we have not checked this.} In particular, the quadratic 
divergences in this expression,
\[
 \frac{\kappa}{2} \hbar g^2 \mu^\epsilon \delta_{\mu\nu} \int \frac{d^d k}
 {(2\pi)^d} \left[\frac{1}{(k+p)^2} + \frac{1}{(k-p)^2} - \frac{2}{k^2}\right]
\]
cancel in {\em any} regularisation scheme that allows to shift the
loop momentum.

Putting all results together, we obtain for the divergent part of the 
one-loop photon vacuum polarisation
\begin{equation}
\neu{\Pi_{\mu\nu}}(p) = - \hbar g^2 \frac{11 \kappa}{24 \pi^2 \epsilon} \left[
\delta_{\mu\nu} p^2 - p_\mu p_\nu\right] + {\cal O}(\hbar ^2).
\label{<107>}
\end{equation}

\subsection{The One-Loop Beta-Function}
\label{sec:3.4}
The counterterm for (\ref{<107>}) in the MS-scheme
is fixed by the requirement that it cancels the divergence in 
$\neu{\Pi}$ without any finite part. In general, we will express the action
in terms of the renormalised quantities and add the counterterm contribution 
to obtain finite Green's functions. Alternatively, the Lagrangian can be 
written 
in terms of the bare fields $\left\{a_\mu^{(0)},\bar{\eta}_0,\eta_0\right\}$
and couplings $g_0$, where $g_0$ is not running and the residue of the
propagator for the bare fields is fixed to 1 by the canonical commutation
relations. For the Abelian vacuum polarisation, this results 
in\footnote{Since the vacuum polarisation is transversal, there is no 
correction to the longitudinal part of $D_0^{-1}$. Thus, the gauge fixing
parameter $\alpha$ is not renormalised.}   
\begin{eqnarray}
\Gamma+\delta \Gamma &=& \frac{1}{2 g^2 \mu^\epsilon} \int \frac{d^4 p}
{(2\pi)^4}\,a_\mu(-p) \left[\delta_{\mu\nu} p^2 - p_\mu p_\nu\right]\,a_\nu(p) 
\cdot \left(1+\hbar g^2 \frac{11 \kappa}{24 \pi^2 \epsilon}\right) + \cdots
\nonumber \\
{}&\stackrel{!}{=}& \frac{1}{2 g_0^2}\int \frac{d^4 p}{(2 \pi)^4}\,
a^{(0)}_\mu(-p) \left[D_0^{-1}\right]_{\mu\nu}(p)\,a^{(0)}_\nu(p) + \cdots
\label{this}
\end{eqnarray}
and our definitions (\ref{Z}) lead to
\begin{equation}
Z_n = 1 + \hbar g^2 \frac{11\kappa}{24 \pi^2\epsilon} + \cdots .
\end{equation}
The coupling constant renormalisation is given by the loop corrections to 
any of the vertices of the theory. For the $\bar{\eta}\eta a_\mu$-vertex,
this results in the relation (\ref{Z}) between the bare and 
renormalised couplings
\begin{equation}
g_0 = g \mu^{\epsilon/2}\cdot Z_g\,Z_n^{-1/2}\,Z_{\rm gh}^{-1}.
\end{equation}

As explained in section \ref{sec:3.2}, the cancellation 
$Z_g = Z_{\rm gh}$ holds, so that the coupling
constant from the $\eta\bar{\eta}a_\mu$-vertex is determined by the
Abelian vacuum polarisation (\ref{<107>}) alone:
\begin{equation}
g_0 = g \mu^{\epsilon/2} Z_n^{-1/2}  
= g \mu^{\epsilon/2} \left[1 - \hbar g^2 \frac{11 \kappa}{48 \pi^2 \epsilon}
 + {\cal O}(\hbar^2) \right]. 
\label{<108>}
\end{equation}
By gauge invariance, a similar cancellation must be present in any other
definition of the coupling constant (involving $a_\mu$), e.g.~the 
triple gluon vertex. Note that in this respect, the {\sf MAG}
closely resembles the background gauge formalism, with the neutral photon
figuring as background field.

From (\ref{<108>}) the $\beta$-function is easily found as $\epsilon \to 0$:
\begin{equation}
\beta(g) \equiv \mu \frac{\partial g}{\partial\mu} = 
- \hbar\cdot\beta_0 g^3 + {\cal O}(\hbar^2)\qquad ; \qquad
\beta_0 = \frac{11 \kappa}{48 \pi^2}  > 0.
\label{<109>}
\end{equation}
This is the known one-loop result for a gauge group $G=SU(2)$. 

In summary, we have shown that the semiclassical expansion around the trivial 
saddle point of our reformulated theory coincides with the standard 
perturbation theory. In particular, we find that the vacuum polarisation of 
the Abelian gauge field alone accounts for the asymptotic freedom of  $SU(2)$ 
YM theory in the {\sf MAG}. The charged field contributions cancel exactly.

\section{ Conclusions } 
\label{sec:4}
In this paper we have presented a reformulation of $SU(2)$ Yang-Mills theory
in the maximal Abelian gauge where the charged gauge fields $\ch{A_\mu}$ 
are exactly integrated out at the expense of an Abelian tensor field 
$\chi_{\mu\nu}$. From a physical standpoint, this 
field (or at least its saddle point value) describes the commutator term of 
the YM field strength in a non-perturbative manner. 

The upshot of this formulation is an effective Abelian theory equivalent
to $SU(2)$ YM with monopole-like gauge fixing singularities. The action
contains a free Maxwell piece and a manifest $U(1)$ invariant 
self-interaction $S_{\rm int}$ described in terms of the new field 
$\chi$ (and the FP ghosts for the {\sf MAG}). We derived the equation of 
motion for these fields and showed that, unlike the so-called field strength
approach \cite{R9}, there exists a trivial solution with a definite physical 
meaning: It gives access to the standard loop expansion, i.e.~the short 
distance physics. 

The detailed analysis of the semiclassical expansion around this trivial
saddle point shows that the Abelian self-interaction alone is responsible
for the asymptotic freedom of the full YM theory. This is due to a 
complete cancellation of all diagrams with external charged legs in the
divergent part of the one-loop vertex correction. The coupling constant
is only renormalised through the Abelian vacuum polarisation, which leads
to the correct value for the one-loop $\beta$-function. In this sense, 
asymptotic freedom in the maximal Abelian gauge can be understood in terms
of the Abelian gauge field (photon) alone.

This important result is another attractive feature of {\sf MAG}. It shows that
in this gauge, not only the low-energy confinement properties (string tension)
is dominated by the Abelian field configurations (as seen on the lattice
\cite{R4}), but also the asymptotic freedom. This fact make this gauge
indeed very attractive.  

In the non-perturbative regime, recent lattice calculations indicate that the 
long distance behaviour of YM theory in {\sf MAG} comes almost entirely from 
the Abelian field configurations, i.e.~the effective Abelian theory 
gives rise to absolute colour confinement by a condensation of 
the aforementioned magnetic monopoles leading to the dual Meissner effect. 
As yet, we do not have a satisfying theoretical understanding for the 
suppression of charged gluon dynamics in low energy observables like the Wilson 
loop (Abelian dominance). The modification of the charged gluon propagator 
caused by non-trivial solutions of the $\chi$-eom could provide further insight
into this question, as will be shown in a forthcoming publication.

\appendix
\section{The Propagators $\DD$ and $\LL$}
\label{app:1}
In this appendix we give the full form of the multiplier and charged gluon
propagator. This illustrates the shorthand definition given in the main text.
Firstly, $\LL^{-1}$ can be interpreted as propagator of the
multiplier field $\phi$. Its inverse is given explicitly by 
\begin{equation}
\LL_{\bar{a}\bar{b}}(x,y) := - \neu{\hat{D}_{\mu,\bar{a}\bar{c}}}(x) 
\left[\HH^{-1}\right]^{\mu\nu}_{\bar{c}\bar{d}}(x,y)
\neu{\hat{D}_{\nu,\bar{d}\bar{b}}}(y) \equiv - \left[\neu{\hat{D}}
\HH^{-1}\neu{\hat{D}}\right]_{\bar{a}\bar{b}}.
\end{equation}
The operator $\DD$, on the other hand, has been identified in the main text
as the charged gluon propagator in {\sf MAG}. The precise meaning of 
formula (\ref{<28>}) and the subsequent definition of the projectors 
$\TT$, $\TT^\dagger$ can be taken from the detailed form
\begin{eqnarray}
\DD^{\mu\nu}_{\bar{a}\bar{b}}(x,y) := 
\left[\HH^{-1}\right]^{\mu\nu}_{\bar{a}\bar{b}}(x,y) 
&+&\int\left[\HH^{-1}\right]^{\mu\alpha}_{\bar{a}\bar{c}}
(x,x_1)\,\neu{D_{\alpha,\bar{c}\bar{d}}}(x_1)\,
\LL^{-1}_{\bar{d}\bar{e}}(x_1,x_2)\cdot\nonumber\\ 
&\cdot&\neu{D_{\beta,\bar{e}\bar{f}}}(x_2)
\,\left[\HH^{-1}\right]^{\beta\nu}_{\bar{f}\bar{b}}
(x_2,y)\,\,d^4x_1\,d^4x_2 .
\end{eqnarray}

\section{The Legendre Transformation}
\label{app:2}
Using the superfield formalism of section \ref{sec:3.1}, we start with
the YM generating functional in the form (\ref{<200>}), i.e.
\begin{equation}
Z[\theta,\lambda] = \int {\cal D}\Theta\exp\left\{-\frac{1}{\hbar g^2}
\left({\cal F} + {\cal F}_{\rm cur}\right)\right\}.
\label{a2.1}
\end{equation}
The part ${\cal F}$ of the exponent independent of the currents has the 
detailed form
\begin{equation}
{\cal F} = \frac{1}{2} \int a_\mu [D_0^{-1}]_{\mu\nu} a_\nu + \frac{1}{4} 
\int\chi_{\mu\nu} \chi_{\mu\nu} + \int \bar{\eta} \left(- \neu{D_\sigma} 
\neu{D_\sigma}\right) \eta + \frac{1}{2} \Tr\ln \HH + \frac{1}{2}\Tr\ln \LL
\,.
\end{equation} 
Furthermore, using the bracket notation to indicate summation or integration 
over relevant indices, the current part of the action can be rewritten in 
components as
\begin{eqnarray}
- {\cal F}_{\rm cur}[\Theta,\theta,\lambda] &=& \left<\neu{j}\cdot\neu{a}
\right> + \left<\bar{\xi}\cdot\eta + \bar{\eta}\cdot\xi\right> + 
\left<\vartheta\cdot\chi\right> + \frac{1}{2}\left<\ch{j}\cdot\DD\cdot
\ch{j}\right> + 
\nonumber \\
& &{}+ \frac{1}{2}\left<\varphi\cdot\LL^{-1}\cdot\varphi\right>
+ i \left<\varphi\cdot\LL^{-1}\neu{D}\HH^{-1}\cdot\ch{j}\right> .
\end{eqnarray}
The calculation of (\ref{a2.1}) in a semiclassical expansion and the 
definition of the Legendre transformation to the effective action is 
given in the main text, see section \ref{sec:3.1}. As indicated in formula
(\ref{<204>}), we should eliminate the currents in $\Gamma$ in favour of 
the classical fields. Usually, however, one does not need to do this 
explicitly to one loop order in $\Gamma$: 
\begin{itemize}
\item The implicit dependence on the currents through the saddle point
      $\bar{\Theta}$ drops out, since $\bar{\Theta}$ and the classical 
      field $\vek{\Theta}$ differ by a term of order ${\cal O}(\hbar)$ which
      leads to a ${\cal O}(\hbar^2)$ effect in the effective action 
      $\Gamma$.
      We can simply replace $\bar{\Theta}\rightarrow\vek{\Theta}$.
\item The explicit dependence of $\Gamma$ on the currents cancels 
      between the term ${\cal F}_{\rm cur}$ in the tree level action 
      and the linear current term in the definition (\ref{<204>}) of the 
      effective action. This is only true for the currents $\theta$ 
      corresponding to fields that have not yet been integrated out. Clearly, 
      the exact integration over $\Lambda = \{\ch{A_\mu},\phi\}$ in our 
      approach leads to {\em quadratic} current terms in ${\cal F}_{\rm cur}$ 
      and thus, the currents $\lambda = \{\ch{j_\mu},\varphi\}$ do 
      {\em not} drop out of the effective action.
\end{itemize}  
In superfield notation, the first two terms in the loop expansion of the 
effective action are given by
\begin{eqnarray}
g^2\Gamma_0[\vek{\Theta},\vek{\Lambda}] &=& {\sf F}[\Theta,\lambda] + 
\int \vek{\Lambda}\lambda 
\label{a2.2}
\\
g^2\Gamma_1[\vek{\Theta},\vek{\Lambda}] &=& \frac{g^2}{2}\Tr\ln\HH + 
\frac{g^2}{2}\Tr\ln\LL + \frac{g^2}{2}\Tr\ln\left.\frac{\delta^2 {\sf F}
[\Theta,\lambda]}{\delta\Theta^2}\right|_{\vek{\Theta}} .\nonumber
\end{eqnarray}
As mentioned above, the operator ${\sf F}$ in this equation still contains the
currents $\lambda = \{\ch{j_\mu},\varphi\}$ quadratically:
\begin{eqnarray}
{\sf F} &=& \frac{1}{2}\left<a\cdot D_0^{-1}\cdot a\right> + \frac{1}{4}
\left<\chi\cdot\chi\right> + \left<\bar{\eta}\cdot\left(-\neu{D}\neu{D}\right)
\cdot\eta\right> -
\label{a2.3} \\
&& {} - \frac{1}{2}\left<\ch{j}\cdot\DD\cdot\ch{j}\right> - \frac{1}{2}
\left<\varphi\cdot\LL^{-1}\cdot\varphi\right> - i \left<\varphi\cdot\LL^{-1}
\neu{D}\HH^{-1}\cdot\ch{j}\right> . 
\nonumber
\end{eqnarray}
Our final task is now to eliminate $\lambda= \{\ch{j_\mu},
\varphi\}$ in this expression
in favour of the classical fields $\vek{\Lambda} = \{{\sf A}_\mu,\vek{\phi}\}$
by inverting the definition (\ref{<203>}). Once again, this only needs to 
be done to lowest order. It is useful to work in components rather than 
superfields, with the result 
\begin{eqnarray}
\varphi &=& - i\,\,\neu{D_\mu} \ch{{\sf A}_\mu} 
\nonumber \\
\TT\cdot\ch{j} &=& \HH\cdot \TT^{\dagger}\cdot\ch{A} = \TT\cdot\HH\cdot\ch{A} 
\nonumber \\
\neu{D_\mu}\,\HH^{-1}_{\mu\nu}\cdot\ch{j_\nu} &=& (-i)\,\,\LL\cdot \vek{\phi} + 
\neu{D_\mu} \ch{{\sf A}_\mu}.
\label{a2.4}
\end{eqnarray} 
These equations must be used to eliminate $\ch{j_\mu}$ and $\varphi$
from the operator ${\sf F}$ (leading to rather messy expressions). 

Note also that the equations (\ref{a2.4}) can be used to eliminate
the classical field $\vek{\chi}$ from the effective action, as explained in
the main text. The calculation is straightforward, though lengthy: 
We replace $\vek{\chi}$ by the saddle
point value $\bar{\chi}$ , which is allowed at one-loop order. Since
$\bar{\chi}$ depends on the currents $\lambda$, we can directly use 
equations (\ref{a2.4}) to obtain the result (\ref{<207>}) quoted in the 
main text.  

Let us close this section by considering the special case when the 
classical fields $\{{\sf A}_\mu,\vek{\phi}\}$ and thus the currents 
$\lambda$ vanish. The operator ${\sf F}$ simplifies to the first line in 
(\ref{a2.3}) and the supertrace in the one-loop correction $\Gamma_1$ 
from (\ref{a2.2}) can actually be decomposed in its component blocks.
One finds the result (\ref{<208>}) quoted in the main text, where the
explicit form of the abbreviation $\Delta_{\mu\nu}$ reads 
\begin{eqnarray}
\Delta_{\mu\nu}(x,y) &=& \left\{2 \bar{\vek{\eta}}^{\bar{a}} {\sf a}_\mu + 
\epsilon^{\bar{a}\bar{c}3} \left[\left(\partial_\mu^x
\bar{\vek{\eta}}^{\bar{c}}(x)\right) - \bar{\vek{\eta}}^{\bar{c}}(x) 
\partial_\mu^x\right]\right\}\, \left[-\neu{D_\sigma}\neu{D_\sigma}
\right]^{-1}_{\bar{a}\bar{b}}(x,y) \cdot
\nonumber \\
&& {} \cdot \left\{2 \vek{\eta}^{\bar{b}}{\sf a}_\nu + 
\epsilon^{\bar{b}\bar{d}3}
\left[\left(\partial_\nu^y\vek{\eta}^{\bar{d}}(y)\right) - 
\stackrel{\leftarrow}{\partial_\nu^{y}} \vek{\eta}^{\bar{d}}(y)\right]
\right\}.
\end{eqnarray}

\section{Proper Time Regularisation of Loop Integrals}
\label{app:3}
The basic idea of the proper time method is to represent the typical 
propagators by a parameter integral and then to cut off the parameter
(the {\em proper time}\/) instead of the loop integration itself: 
\begin{eqnarray*}
\frac{1}{k^2+m^2} &=& \int\limits_0^\infty ds\,e^{-s(k^2+m^2)}
\longrightarrow \int\limits_{1/\Lambda^2}^\infty ds\,e^{-s(k^2+m^2)}
\\
\frac{1}{(k^2+m^2)^2} &=& \int\limits_0^\infty ds\,s\,e^{-s(k^2+m^2)}
\longrightarrow \int\limits_{1/\Lambda^2}^\infty ds\,s\,e^{-s(k^2+m^2)}.
\end{eqnarray*}
Here, $\Lambda$ corresponds to a UV momentum cutoff. The obvious advantage
of this method is that the loop integration itself is left unchanged and we can 
perform all the standard manipulations on it. In particular we are allowed 
to {\em shift} the loop momentum. Examples are
\begin{eqnarray}
\int \frac{d^4 k}{(2 \pi)^4} \frac{1}{k^2} &\longrightarrow& 
\frac{\Lambda^2}{16 \pi^2} + \cdots 
\nonumber \\
\int \frac{d^4 k}{(2 \pi)^4}\frac{k_\mu k_\nu}{k^2 (k+p)^2}
&\longrightarrow& \frac{\delta_{\mu\nu} p^2 - 4 p_\mu p_\nu}{192 \pi^2}
\ln \frac{p^2}{\Lambda^2} + \frac{\delta_{\mu\nu} \Lambda^2}{32 \pi^2}
+\cdots
\\
\int \frac{d^4 k}{(2 \pi)^4}\frac{1}{k^2 (k+p)^2} &\longrightarrow&
 - \frac{1}{16 \pi^2}\ln\frac{p^2}{\Lambda^2} + \cdots .
\nonumber
\end{eqnarray}
The omitted terms are finite as $\Lambda\to\infty$. The first two examples
clearly show that the proper time method is able to parametrise quadratic 
divergences in a gauge invariant way. 

As an example of the cancellation of such divergences consider the 
$\HH$ contribution to the Abelian vacuum polarisation. The relevant term
$\Gamma_1 \sim \frac{1}{2} \Tr\ln \HH$ of the effective action is 
separately invariant under the residual $U(1)$ symmetry (\ref{<24a>}), 
provided the regularisation of the trace respects this. As a
consequence, we expect the contribution to the Abelian vacuum polarisation
to be transversal and the dangerous quadratic divergences should cancel 
exactly. 

Let us now check this expectation using the proper time method. The 
contributing diagrams are (see (\ref{<105>}) in the main text)
\[
\begin{minipage}{6.8cm}
\epsfig{file = vacpol2.ps,width = 6.5cm}
\end{minipage}. 
\]
In momentum space, the first diagram reads
\begin{equation}
\hbar g^2 \kappa\,\delta_{\mu\nu}\int \frac{d^4 k}{(2\pi)^4} \frac{4}{k^2}
= \hbar g^2 \kappa\,\delta_{\mu\nu} \frac{\Lambda^2}{4 \pi^2} + \cdots.
\end{equation}
We observe that the proper time method really {\em parametrises} the quadratic
divergence, whereas this integral would be simply zero in dimensional
regularisation.

The second diagram above is given by 
\begin{eqnarray}
&-& \hbar g^2 \frac{\kappa}{2}
\int \frac{d^4 k}{(2\pi)^4} \frac{8\, \delta_{\mu\nu} p^2 + 16 k_\mu k_\nu
+ 8 (k_\mu p_\nu + k_\nu p_\mu) - 4 p_\mu p_\nu}{k^2 (k+p)^2} = 
\nonumber \\
& & = \hbar g^2 \kappa\cdot 5\,\frac{\delta_{\mu\nu} p^2 - p_\mu p_\nu}
{24 \pi^2} \ln\frac{p^2}{\Lambda^2} - \hbar g^2 \kappa\cdot
\frac{\delta_{\mu\nu} \Lambda^2}{4 \pi^2} + \cdots.
\end{eqnarray}
Clearly the quadratic divergences cancel {\em exactly} between the 
two diagrams and the remaining logarithmic divergence has the required
transversal form. On comparing with the result of the main text, where we
used dimensional regularisation, we observe the correspondence
$1/\epsilon \leftrightarrow 1/2\ln(\Lambda^2/p^2)$ between the two 
procedures. 

In all the other diagrams containing quadratic divergences, the calculation
(and the correct cancellation) is completely equivalent to the above 
example.

\end{document}